\documentclass[preprint,showpacs,preprintnumbers,amsmath,amssymb,nofootinbib]{revtex4}

\usepackage{graphicx}
\usepackage{epstopdf}
\usepackage{axodraw}
\usepackage{dcolumn}
\usepackage{bm}

\newcommand{\nn}{\nonumber}

\newcommand{\be}{\begin{eqnarray}}
\newcommand{\ee}{\end{eqnarray}}
\catcode`\@=11
\def\lsim{\mathrel{\mathpalette\@versim<}}
\def\gsim{\mathrel{\mathpalette\@versim>}}
\def\@versim#1#2{\vcenter{\offinterlineskip
\ialign{$\m@th#1\hfil##\hfil$\crcr#2\crcr\sim\crcr } }}
\catcode`\@=12

\begin{document}
\vspace{2cm}
\preprint{KANAZAWA-07-18}

\title{
Flavor Changing Neutral Higgs Bosons 
in a Supersymmetric Extension
 based on a $Q_6$ Family Symmetry}

\author{Naoko Kifune$^a$}
\author{Jisuke Kubo$^a$}
\author{Alexander Lenz$^b$}

\affiliation{
${}^a$ Institute for Theoretical Physics, Kanazawa
University, Kanazawa 920-1192, Japan\\
${}^b$
Universit\"at Regensburg, D-93051 Regensburg, Germany
\vspace{3cm}
}

\begin{abstract}

\vspace{1cm} 

A supersymmetric extension of the standard model
 based on the discrete $Q_6$ family symmetry is considered,
 and we investigate flavor-changing neutral current (FCNC)
 processes, especially those mediated 
 by heavy flavor-changing neutral Higgs bosons.
 Because  of the family symmetry
 the number of the independent Yukawa couplings is smaller
 than that of the observed quantities such as
 the Cabibbo-Kobayashi-Maskawa matrix
 and the quark masses, so that
 the FCNCs can be parametrized only by
 the mixing angles and masses of  the Higgs fields.
 We focus our attention on the mass differences of the neutral
 $K, D$ and $B$ mesons.
All the constraints 
including  that from the ratio $\Delta M_{B_s}/\Delta M_{B_d}$ can be satisfied, 
if the heavy Higgs bosons are heavier
than $\sim 1.5$ TeV.
If the constraint from 
$\Delta M_K$ is slightly relaxed,
the heavy Higgs bosons can be as light as
$\sim 0.4$ TeV, which is
 within the accessible range of LHC.
 
 \end{abstract}

\maketitle

\section{Introduction}
In recent studies on flavor symmetries
\footnote{For recent reviews see, for instance, 
 \cite{Altarelli:2007gb,Ma:2007ia,Mondragon:2006hi}.}
it has become clear that  a
 flavor symmetry can be realized at low energies.
 As long as this possibility is not excluded,
theoretical as well as experimental searches for 
 a low energy flavor symmetry should be continued.
An important prediction 
 of  any viable low energy flavor symmetry, which is  broken
 only spontaneously or at most softly,
 is the existence of multiple $SU(2)_L$ 
 doublet Higgs fields, as one could 
read off   from  a sort of no-go theorem of \cite{Koide:2004rd}.
 This implies that there should exist
 several neutral Higgs fields that have flavor changing couplings
 to the fermions at the tree-level.
 Therefore, an observation of a non-standard
 flavor changing neutral current (FCNC) process, at LHC for instance,
 is not necessarily an indication of supersymmetry 
 \cite{Buttar:2006zd,Marinelli:2007bk}.
 
 In  Ref.~\cite{Babu:2004tn}
  a supersymmetric flavor model 
based on a dicyclic dihedral group $Q_6$ 
has been suggested. \footnote{
$Q_6$ is one of $Q_{2N}$ with 
$N=2,3,\dots$, which are the ``covering
groups'' of the dihedral groups
$D_N$ \cite{Frampton:1994rk,Frampton:2000mq}.
In recent years there are a number of interesting flavor models
based on $Q_{2N}$ and $D_N$. 
For instance, $D_4$ has been used as a flavor symmetry 
in Refs. 
\cite{Seidl:2003wy,Balaji:2003st,Grimus:2003kq,Grimus:2004rj,Hagedorn:2005kz},
while $D_5, D_6, D_7$ and $Q_4$ have been considered 
in Refs. \cite{Hagedorn:2006ir}, 
\cite{Kajiyama:2006ww}, \cite{Chen:2005jt}
and \cite{Frigerio:2004jg}, respectively.
See also  Refs.~\cite{Okubo:1975hk,Blum:2007jz}.}
The main motivation there was to derive a modified Fritzsch mass 
matrix for the quarks from a flavor symmetry.
With an assumption that CP is spontaneously broken,
the model can fix six quark masses and four 
Cabibbo-Kobayashi-Maskawa (CKM)
parameters in term of nine  parameters of the model.
It has been later realized in  Refs.~\cite{Kubo:2005ty,Kajiyama:2005rk} that 
through an appropriate change of the lepton assignment 
the leptonic sector can be brought into the same form
as that of the model of   \cite{Kubo:2003iw,Kubo:2003pd}.
Then there are only seven parameters
in the leptonic sector  of the model to 
fix six  lepton masses and six Maki-Nakagawa-Sakata (MNS)
parameters.
The discrete flavor group $Q_6$ is the smallest non-abelian
group with which the above situation can be achieved.

However, it turned out that one has to introduce
a certain set of $SU(2)_L\times U(1)_Y $ singlet fields and also additional
abelian global symmetries to make the model viable.
Nothing is wrong with this situation, but in this paper
we would like to stress the minimal content
of the Higgs fields and at the same time
a ``one + two'' structure for each family;
one $Q_6$ singlet and one $Q_6$ doublet for
each family including the $SU(2)_L$ doublet Higgs fields.
In Sec. II we will shed light upon the relation between
the non-renormalization theorem and flavor symmetry, and will show
that different flavor symmetries  can be consistently introduced
into a softly broken supersymmetric gauge theory.
We will systematically investigate this possibility
in a general framework.
With this observation we will find in Sec. III that
the one + two structure of family in  a minimal $Q_6$ extension of
the supersymmetric standard model (MSSM)
 can be consistently realized.

In Sec. IV we will consider the Higgs sector.
Because of the one + two structure the Higgs sector
is much simpler than that of 
\cite{Babu:2004tn,Kubo:2005ty,Kajiyama:2005rk},
and therefore the sector  can be investigated 
with much less assumptions.
We will explicitly show that it is possible to fine tune the 
soft-supersymmetry-breaking (SSB)  parameters
 so as to make the heavy Higgs bosons much heavier
 (several TeV)  than
$M_Z$ and at the same time to obtain
a desired size of spontaneous CP violation 
to reproduce the Kobayashi-Maskawa CP violating phase.

In Sec. V we will first
calculate the unitary matrices
that diagonalize the fermion mass matrices,
which are needed to write down the Yukawa couplings
in terms of mass eigenstates.
We only briefly mention FCNCs and CP violations in the
SSB sector and in the lepton sector, 
because detailed investigations on these subjects have been  recently
carried out in  Ref.~\cite{Kajiyama:2005rk} 
and in  Ref.~\cite{Mondragon:2007af}, respectively.
Instead we investigate FCNC processes mediated
by neutral heavy Higgs fields.
We concentrate on the constraints coming
from the mass differences in the neutral
meson systems, 
$\Delta M_{K}, 
\Delta M_{B_s},
\Delta M_{B_d}$ and $\Delta M_{D}$,
in a similar spirit as Refs.~
\cite{Gabbiani:1996hi,Ciuchini:2002uv,Silvestrini:2005zb,Ciuchini:2006dx,Ciuchini:2007cw,Golowich:2007ka,Lenz:2007nj}
and references therein.
We  express the relevant  flavor changing-neutral
Yukawa couplings in terms of the mass eigenstates, where
except the phases the size of the Yukawa couplings are
basically fixed.
Allowed ranges in which the constraints are satisfied
are shown in different figures.
We find that
the heavy Higgs bosons should be heavier
than $\sim 1.5$ TeV,
although it is possible to fine tune the parameters
such that the constraints can be satisfied
for lighter mass values.

Sect. VI is devoted for conclusion.

\section{
Non-renormalization theorem and
flavor symmetry}

A flavor symmetry can control the structure of
the independent parameters of a theory.
In  supersymmetric theories, moreover,
the non-renormalization
 theorem allows to suppress
certain  couplings
and also to  relate them with each other, 
without facing contradictions with renormalization.
What is therefore the (technical) role of  a flavor symmetry in 
supersymmetric theories?
We  recall that the D-terms are
renormalized and the wave function renormalization
can mix matter superfields $\Phi_i$'s in general.
Therefore, starting with  diagonal kinetic terms
$\Phi_i^*\Phi_i$ is not always consistent with renormalization.
If a non-diagonal (infinite) kinetic term is induced,
a corresponding non-diagonal counter term should be added.
Then after the diagonalization 
the originally assumed structure of the couplings
in the superpotential
will receive  large quantum corrections.
In other words, we have in spite 
of the non-renormalization theorem more parameters
in the superpotential,
when written in terms of the bare fields, than originally assumed.
The undesired mixing among $\Phi_i$'s and 
 large quantum corrections can be avoided 
 if an appropriate  flavor symmetry is present.

We will see below that
the non-renormalization theorem and the renormalization
properties of the soft-supersymmetry-breaking (SSB) terms
allow us to introduce  in a consistent manner different
flavor symmetries for  different sectors of a softly
broken supersymmetric theory to
control the independent parameters  of the theory.

To be more specific, we consider
an $N=1$
supersymmetric gauge theory whose superpotential
is given by
\be
W(\Phi) &= &W_Y(\Phi)+W_\mu(\Phi)~,
\ee
with
\be
W_Y(\Phi) &= &\frac{1}{6} Y^{ijk} \Phi_i \Phi_j \Phi_k~\mbox{and}~
W_\mu(\Phi) =\frac{1}{2} \mu^{ij} \Phi_i\Phi_j ~.
\ee
The SSB Lagrangian can be written as
\be
L(\Phi,W) &=& - \left( ~\int d^2\theta\eta (  \frac{1}{6} 
 h^{ijk} \Phi_i \Phi_j \Phi_k +  \frac{1}{2}  b^{ij} \Phi_i \Phi_j 
+  \frac{1}{2}  M_gW_A^\alpha W_{A\alpha} )+
\mbox{h.c.}~\right)\nn\\
& &-\int d^4\theta\tilde{\eta} \eta \overline{\Phi^j}                   
(m^2)^i_j(e^{2gV})_i^k \Phi_k~,
\ee
where $\eta = \theta^2$, 
$\tilde{\eta} = \tilde{\theta}^2$ are the external
spurion superfields and $M_g$ is the gaugino mass.
The $\beta$ functions of the $Y, \mu, h$ and $m^2$
  are given by  Refs.~\cite{Yamada:1994id}-\cite{Kraus:2002uh}
\be
\beta_Y^{ijk}&=&\gamma^i{}_l Y^{ljk}+\gamma^j{}_lY^{ilk}
+\gamma^k{}_l Y^{ijl}~,
\label{betaY}\\
\beta_\mu^{ij}&=&\gamma^i{}_l \mu^{lj}+\gamma^j{}_l\mu^{il}~,
\label{betamu}\\
\beta_h^{ijk}&=&\gamma^i{}_lh^{ljk}+\gamma^j{}_lh^{ilk}
+\gamma^k{}_lh^{ijl}-2\gamma_1^i{}_lY^{ljk}
-2\gamma_1^j{}_lY^{ilk}-2\gamma_1^k{}_lY^{ijl}~,
\label{betah}\\
\beta_b^{ij}&=&\gamma^i{}_lb^{lj}+\gamma^j{}_lb^{il}
-2\gamma_1^i{}_l\mu^{lj}-2\gamma_1^j{}_l\mu^{il}~,
\label{betab}\\
(\beta_{m^2})^i{}_j &=&\left[ \Delta 
+ X \frac{\partial}{\partial g}\right]\gamma^i{}_j~,
\label{betam2}\\
{\cal O} &=&\left(M_g g^2{\partial\over{\partial g^2}}
-h^{lmn}{\partial
\over{\partial Y^{lmn}}}\right)~,
\label{diffo}\\
\Delta &=& 2{\cal O}{\cal O}^* +2|M_g|^2 g^2{\partial
\over{\partial g^2}} +\tilde{Y}_{lmn}
{\partial\over{\partial
Y_{lmn}}} +\tilde{Y}^{lmn}{\partial\over{\partial Y^{lmn}}}~,
\label{delta}\ee
where $(\gamma_1)^i{}_j={\cal O}\gamma^i{}_j$, 
$Y_{lmn} = (Y^{lmn})^*$, and 
\be
\tilde{Y}^{ijk}&=&
(m^2)^i{}_lY^{ljk}+(m^2)^j{}_lY^{ilk}+(m^2)^k{}_lY^{ijl}~,\\
X &=&
\frac{-|M_g|^2 C(G)+\sum_l m_l^2 T(R_l) }{C(G)-8\pi^2/g^2}~.
\label{xtilde2}
\ee
Here $X$ of  (\ref{xtilde2}) is the expression
in the renormalization scheme of Novikov {\em et al.} \cite{Novikov:1983uc},
$T(R_l)$ is the Dynkin index of $R_l$, and $C_{2}(G)$
is the quadratic Casimir of the adjoint representation of the
gauge group $G$.
From Eqs.~(\ref{betaY})-(\ref{xtilde2})
we now derive the hierarchical structure of the renormalization
properties of the theory, which is basically the
Symanzik theorem applied to  softly broken
supersymmetric gauge theories:
\begin{enumerate}
\item
The (infinite) renormalization of the supersymmetric
parameters $Y^{ijk}, \mu^{ij}$ is not influenced by the 
SSB terms, in accord with the definition of the SSB terms.

\item
The (infinite) renormalization of  the tri-linear
couplings $h^{ijk}$ does not depend
 on $\mu^{ij}$.
 It is also independent on
 $(m^2)^i_j$ and $ b^{ij}$.

\item
The (infinite) renormalization of  the
soft scalar masses $(m^2)^i_j$
 does not depend on $b^{ij}$ and  $\mu^{ij}$,
 as one can see from  Eqs.~(\ref{betam2})- (\ref{xtilde2}).
 
 \item
The (infinite) renormalization of   $b^{ij}$
 does not depend $(m^2)^i_j$ and  $h^{ijk}$,
 which is the consequence of (\ref{betab}).

\end{enumerate}
Because of these renormalization
properties we can
consistently  introduce
different symmetries for different sectors.

To begin with  we assume the existence of a flavor symmetry
in the Yukawa sector which protects the mixing 
(of the wave function renormalization) among the matter superfields $\Phi_i$'s. \footnote{We also assume that the flavor symmetry is
not gauged.}
This implies that the anomalous dimensions $\gamma^i{}_j$
are diagonal, i.e.,
\be
\gamma^i{}_j &=&\delta^i{}_j~\gamma_j~.
\label{gamma}
\ee
Then Eqs.~(\ref{betaY}) -(\ref{betam2}) become
\be
\beta_Y^{ijk}&=&Y^{ijk}(\gamma_i+\gamma_j+\gamma_k)~,~
\beta_\mu^{ij}=\mu^{ij}(\gamma_i+\gamma_j)~,
\label{betamu3}\\
\beta_h^{ijk}&=&(h^{ijk}-2Y^{ijk}{\cal O}) 
(\gamma_i+\gamma_j+\gamma_k)~,~
\beta_b^{ij}=(b^{ij}-2\mu^{ij}{\cal O})(\gamma_i+\gamma_j)~,\\
\label{betab3}
(\beta_{m^2})_l &=&\left[ \Delta 
+ X \frac{\partial}{\partial g}\right]\gamma_l~,
\label{betam23}
\ee
with $\tilde{Y}^{ijk}=
Y^{ijk}(m^2_i+m^2_j+m^2_k)$.
From these equations we observe:

\vspace{2mm}
\noindent
(a)  The $\mu$ sector can have a 
flavor symmetry which is different from
the flavor symmetry of the Yukawa sector
if both symmetries are compatible 
with respect to renormalization
of $\mu_{ij}$.

\vspace{2mm}
\noindent
(b)  It is consistent to
introduce into the tri-linear couplings the same flavor symmetry
as that of the Yukawa couplings, even if it
is violated in other sectors.

\vspace{2mm}
\noindent
(c)  The flavor symmetry which protects the mixing among $\Phi_i$'s
ensures that $(m^2)^i_j$ is diagonal.
If  the Yukawa couplings
and  tri-linear couplings have the
flavor symmetry,
the soft scalar mass terms, too,  can have
the flavor symmetry,
even if the $\mu$ and $b$ terms
do not respect the flavor symmetry.

\vspace{2mm}
\noindent
(d)   The $b$ terms associated with the $\mu$ terms
should  always exist (see (\ref{betab3})).
But the $b$ sector
has no influence on the infinite renormalization
of the parameters in other sectors.
So the violation of a  symmetry in the $b$ sector 
is absolutely soft.

\vspace{3mm}

In the next section we reconsider the supersymmetric flavor
model of \cite{Babu:2004tn,Kubo:2005ty,Kajiyama:2005rk}
along the line of thought  about a flavor symmetry
in this section.

\section{The model}

The supersymmetric flavor model of 
\cite{Babu:2004tn,Kubo:2005ty,Kajiyama:2005rk}
is based on a dicyclic dihedral group $Q_6$.
If CP is spontaneously broken,
the  nine parameters of the model express six quark masses and four 
CKM parameters. 
In the leptonic sector there are only seven parameters
to  fix six lepton masses and six MNS parameters.
As we announced in the introduction
we would like to stress 
the one + two structure for each family;
a $Q_6$ singlet and a $Q_6$ doublet for
each family including the $SU(2)_L$ doublet Higgs fields.

\subsection{The Yukawa sector}
As in the original model of \cite{Babu:2004tn,Kubo:2005ty,Kajiyama:2005rk}
 we assume that the flavor symmetry of the Yukawa sector
is based on $Q_6$.
In Table \ref{assignment} we write the $Q_{6}$ assignment of the quark, 
lepton and Higgs chiral supermultiplets,
\footnote{The same model exists for $Q_{2N}$
if $N$ is odd and a multiple  of $3$.} where
$Q, Q_3, L, L_3$ and  $ H^u, H_3^u, H^d, H_3^d$
stand for $SU(2)_L$ doublet
supermultiplets for the quarks, leptons and Higgs bosons, respectively.
Similarly, $SU(2)_L$ singlet
supermultiplets for quarks, charged leptons and neutrinos are denoted by
$U^c, U^c_3,D^c, D^c_3, E^c, E^c_3$ and $N^c, N^c_3$.
From Table \ref{assignment}
 we see that the one+two  structure of  family is realized,
and because of this structure the $Q_6$
flavor symmetry  can ensure that
no non-diagonal kinetic term can be induced.
So (\ref{gamma}) is satisfied.

\vspace{0.5cm}
\begin{table}
\begin{center}
\begin{tabular}{|c|c|c|c|c|c|c|c|c|c|c|c|}
\hline
 & $Q$ 
 & $Q_3$  
& $U^c,D^c$  
& $U^c_3,D^c_3$ 
 & $L$ & $L_3$ 
 &$E^c,N^c$ & $E_3^c$  &   $N_3^c$ 
  & $H^u,H^d$
 & $H^u_3,H^d_3$ 
\\ \hline
$Q_6$ &${\bf 2}_1$ & ${\bf 1}_{+,2}$ &
 ${\bf 2}_{2}$ &${\bf 1}_{-,1}$ &${\bf 2}_{2}$ &
${\bf 1}_{+,0}$  & ${\bf 2}_{2}$ & 
${\bf 1}_{+,0}$ & ${\bf 1}_{-,3}$ &
${\bf 2}_{2}$ & ${\bf 1}_{-,1}$  \\ \hline
$R$ & $-$ & $-$ &
$-$  & $-$ & 
$-$ & $-$ & $-$ & 
$-$ & $-$  
& $+$ & $+$   \\ \hline
\end{tabular}
\caption{ \footnotesize{The $Q_{6} \times R$ assignment 
of the chiral matter supermultiplets, where $R$ is the $R$ parity.
The group theory notation is given in 
 Ref.~\cite{Babu:2004tn}.}}
\label{assignment}
\end{center}
\end{table}
We then write down the most general, renormalizable,
$Q_{6} \times R$ invariant superpotential $W$ ($R$ is the $R$ parity.):
\be
W_Y &=& W_Q+W_L~,
\ee
where
\be
W_Q
&=&\sum_{I,i,j,k=1,2,3}\left(
Y_{ij}^{uI} Q_{i} U_{j}^c  H^u_I
+Y_{ij}^{dI} Q_{i} D_{j}^c H^d_I\right)~,
\label{wQ}\\
W_L
&=&\sum_{I,i,j,k=1,2,3}\left(
Y_{ij}^{eI} L_{i} E_{j}^c  H^d_I
+Y_{ij}^{\nu I} L_{i} N_{j}^c H^u_I\right)~.
\label{wL}
\ee
The Yukawa matrices $Y$'s are given by
\be
{\bf Y}^{u1(d1)} &=&\left(\begin{array}{ccc}
0 & 0 & 0 \\
0 & 0 & Y_b^{u(d)} \\
0&  Y_{b'}^{u(d)}  & 0 \\
\end{array}\right),~
{\bf Y}^{u2(d2)} =\left(\begin{array}{ccc}
0 & 0 & Y_b^{u(d)}\\
0 & 0 & 0 \\
  -Y_{b'}^{u(d)} &0 & 0 \\
\end{array}\right),\nn\\
{\bf Y}^{u3(d3)}&=&\left(\begin{array}{ccc}
0 & Y_c^{u(d)} & 0\\
Y_c^{u(d)} & 0 & 0 \\
0 &  0 & Y_a^{u(d)} \\
\end{array}\right),
\label{Yuq}
\ee
\be
{\bf Y}^{e1} &=&\left(\begin{array}{ccc}
-Y_c^{e} & 0 & Y_b^{e}\\
0 & Y_c^{e} &  0\\
Y_{b'}^{e}& 0  & 0 \\
\end{array}\right),~
{\bf Y}^{e2} =\left(\begin{array}{ccc}
0 & Y_c^{e} & 0 \\
Y_c^{e} & 0 & Y_b^{e} \\
0&  Y_{b'}^e & 0 \\
\end{array}\right),~
{\bf Y}^{e3}=0,
\label{Yue}
\ee
\be
{\bf Y}^{\nu1} &=&\left(\begin{array}{ccc}
-Y_c^{\nu}& 0 & 0 \\
0 & Y_c^{\nu} & 0 \\
 Y_{b'}^\nu & 0 & 0 \\
\end{array}\right),~
{\bf Y}^{\nu2} =\left(\begin{array}{ccc}
0 & Y_c^{\nu} & 0\\
Y_c^{\nu} &  & 0 \\
0&Y_{b'}^{\nu}  & 0 \\
\end{array}\right),\nn\\
{\bf Y}^{\nu3}&=&\left(\begin{array}{ccc}
0 & 0 & 0\\
0 & 0 & 0 \\
0 &  0 & Y_a^{\nu} \\
\end{array}\right).
\label{Yun}
\ee
All the parameters appearing above are  real, because 
we assume  that   CP is spontaneously
broken.
We will shortly come back to this issue.

\subsection{The $\mu$ sector}

The most general $Q_6\times R$ invariant
renormalizable $\mu$ part of the superpotential is
\be
W_\mu^{(Q_6)}
&=&\mu H^u_I H^d_I +\frac{m}{2} N^c_I N^c_I~.
\label{Wmu1}
\ee
Note that no mass terms for $H^{u,d}_3$ and 
$N^c_3$ are allowed by $Q_6$
and that the superpotential $W_\mu^{(Q_6)}$
has an accidental $O(2)$ symmetry.
For phenomenological reasons we however need mass terms
for $H^{u,d}_3$ and  $N^c_3$. Therefore, we assume that
the flavor symmetry of the $\mu$ sector is 
$O(2)$ and that  $H^{u,d}_3$ and  $N^c_3$
are singlets of $O(2)$, and add
\be
W_\mu^{(Q_6\hspace{-3.5mm}/~)}
&=&\mu_3 H^u_3 H^d_3 +\frac{m_3}{2} N^c_3 N^c_3
\label{Wmu2}
\ee
to (\ref{Wmu1}). 
Then the total $\mu$ part of the superpotential
is $W_\mu=W_\mu^{(Q_6)}
+W_\mu^{(Q_6\hspace{-3.5mm}/~)}.$
The $O(2)\times R$ symmetry of
$W_\mu$ is compatible with $Q_6\times R$ of the Yukawa sector,
because $Q_6$ ensures
\be
\gamma_{H_1^u} &=&\gamma_{H_2^u}~\mbox{and}~
\gamma_{H_1^d}=\gamma_{H_2^d}~.
\ee

\subsection{Soft-supersymmetry-breaking sector}
\subsubsection{The tri-linear couplings and
soft scalar mass terms}
We require that the tri-linear couplings and
soft scalar mass terms have the same flavor symmetry 
as that of the Yukawa sector, that is, $Q_6\times R$.
Therefore, the tri-linear couplings and
soft scalar mass matrices have the following form:
\be
{\bf h}^{k}_{ij} &=&A_{ij}{\bf Y}^{k}_{ij}~,~k=u1,u2,\dots,\nu_3~,
\label{hijk}
\ee
where ${\bf Y}^{k}_{ij}$ are given in (\ref{Yuq})-(\ref{Yun}), and
\be
{\bf m}^2 & \propto &
\left(\begin{array}{ccc}
1 & 0 &0\\
0 & 1 &0\\
0 & 0 & f
\end{array}\right)~.
\label{m2}
\ee
for all the bosonic scalar partners.
This is very crucial to suppress FCNCs
in the SSB sector as we will see later on.

\subsubsection{The $b$ terms}
The $b$ sector should contain
at least terms which correspond to the $\mu$
terms $W_\mu
=W_\mu^{(Q_6)}
+W_\mu^{(Q_6\hspace{-3.5mm}/~)}$,
where $W_\mu^{(Q_6)}$ and
$W_\mu^{(Q_6\hspace{-3.5mm}/~)}$
are given in (\ref{Wmu1}) and (\ref{Wmu2}), respectively, i.e.
\be
{\cal L}_b^{(O_2)}=
&=& b \hat{H}^u_I \hat{H}^d_I +
b_{33} \hat{H}^u_3 \hat{H}^d_3 
+b_N\hat{ N}^c_I \hat{N}^c_I
+b_{N_3} \hat{N}^c_3 \hat{N}^c_3+h.c.
\label{b1}
\ee
(The hatted fields are bosonic components.)
Because of the $O(2)$ symmetry in the $\mu$ and $b$ sectors
and the $Q_6$ symmetry in the soft scalar mass terms,
the Higgs scalar potential  also respects the
$O(2)$ symmetry, so that there is a  Nambu-Goldstone boson
corresponding to this symmetry because 
the $O(2)$ symmetry
the gauge symmetry  is spontaneously broken, together with
$SU(2)_L\times U(1)_Y$.
Moreover,  we  face the domain wall problem
when the discrete flavor symmetries are spontaneously broken.
To overcome these problems we add terms which
explicitly break $O_2$ down to $Z_2$:
\be
{\cal L}_b^{(O_2\hspace{-3.5mm}/~)}=
&=& b_{++} \hat{H}^u_+ \hat{H}^d_+
+b_{--}\hat{H}^u_-\hat{H}^d_-
+b_{+3}\hat{H}^u_+\hat{H}^d_3
+b_{3+}\hat{H}^u_3\hat{H}^d_+\nn\\
& & +b_{N_+}\hat{ N}^c_+  \hat{N}^c_+
+b_{N_-}\hat{ N}^c_- \hat{N}^c_-+
\hat{ N}^c_3\hat{ N}^c_+
+h.c.,
\label{b2}
\ee
where
\be
H^{u,d}_{\pm}&=& \frac{1}{\sqrt{2}}
 (H^{u,d}_1\pm H^{u,d}_2)~,~
  N^c_{\pm}= \frac{1}{\sqrt{2}}
( N_1^c\pm N_2^c)~.
\label{Hpm}
 \ee
($ H^{u,d}_{+}, H^{u,d}_3,
N^c_+$ and $ N^c_3$ are $Z_2$ even, while 
$ H^{u,d}_{-}$ and  
$N^c_-$ are $Z_2$ odd.)
This $Z_2$ is indeed broken by the the Yukawa and 
and tri-linear couplings, but is compatible 
with $Q_6$, i.e.,
$\gamma_{H_1^{u,d}}=
\gamma_{H_2^{u,d}}$.

We allow the $b$ parameters  to be complex,
because CP can not be broken if all the $b$ parameters are real
as we will find in the next subsection.
So CP is explicitly, but only softly broken in this sector.
In Table \ref{symmetry} we give
the symmetry of the each sector.

\vspace{0.5cm}
\begin{table}
\begin{center}
\begin{tabular}{|c|c|c|c|c|}
\hline
 & ${\bf Y},{\bf h}$ & 
 ${\bf m}$
 & $\mu $  sector & $b$ terms
 \\ \hline
$Q_6$ & $\bigcirc$ & $\bigcirc$ & $\times$  &  $\times$
\\ \hline
$O_2$ & $\times$ &$\bigcirc$ & $\bigcirc$  &  $\times$
\\ \hline
$Z_2$ & $\times$ & $\bigcirc$ & $\bigcirc$  &  $\bigcirc$
\\ \hline
CP & $\bigcirc$ & $\bigcirc$ & $\bigcirc$  &  $\times$
\\ \hline
R & $\bigcirc$ & $\bigcirc$ & $\bigcirc$  &  $\bigcirc$
\\ \hline
\end{tabular}
\caption{ \footnotesize{The symmetry of the different sectors.
${\bf Y}, {\bf h}$ and 
 ${\bf m}$ stand for 
 the Yukawa,  tri-linear
 and   soft scalar mass sector, respectively.
 $Q_6$ ensures that all the anomalous dimensions $\gamma$'s
 are diagonal, and that the two components
 of a $Q_6$ doublet have a same anomalous dimension.
 Therefore, $Q_6$ in the Yukawa and tri-linear sectors and
 $O_2$ in the $\mu$ sector are compatible with each other.
 $O_2$ in the soft scalar mass sector is accidental.
$Z_2$ is a subgroup of $O_2$, which implies 
the compatibility of $O_2$ and $Z_2$.
CP is explicitly broken only by the $b$ terms,
which is (super) soft because
the  propagation of its violation
to the other sectors is calculable and small.
So, all the symmetries are compatible with each other.}
 }
\label{symmetry}
\end{center}
\end{table}

\section{The Higgs sector}
\subsection{The Higgs potential}

Given the $O(2)\times R$ 
invariant superpotential $W_\mu$ in the $\mu$
sector (\ref{Wmu1}) and (\ref{Wmu2}) along with the $Q_6\times R$ invariant
soft scalar masses (\ref{m2}) and the $Z_2\times R$ invariant
$b$ terms (\ref{b1}) and (\ref{b2}),
we can now write down the scalar potential.
For simplicity we assume that only the neutral scalar components
 (denoted
by a superscript $0$) of the Higgs supermultiplets acquire VEVs:
\be
V &=&
m_{H_{+}^{u}}^{2}~(|\hat{H}_{+}^{0u}|^{2}+|\hat{H}_{-}^{0u}|^{2})+
m_{H_{+}^{d}}^{2}~(|\hat{H}_{+}^{0d}|^{2}+|\hat{H}_{-}^{0d}|^{2})
+m_{H_{3}^{u}}^{2}~|\hat{H}_{3}^{0u}|^{2}+
m_{H_{3}^{d}}^{2}~|\hat{H}_{3}^{0d}|^{2}
\nn\\
& +&\frac{1}{8}(\frac{3}{5}g_{1}^{2}+
g_{2}^{2})(|\hat{H}_{+}^{0u}|^{2}+
|\hat{H}_{-}^{0u}|^{2} +|\hat{H}_{3}^{0u}|^{2}
-|\hat{H}_{+}^{0d}|^{2}-
|\hat{H}_{-}^{0d}|^{2} -|\hat{H}_{3}^{0d}|^{2} )^{2}\nn\\
& +&\left[ b_{++}' \hat{H}^{0u}_+ \hat{H}^{0d}_+
+b_{--}'\hat{H}^{0u}_-\hat{H}^{0d}_-
+b_{+3}\hat{H}^{0u}_+\hat{H}^{0d}_3
+b_{3+}\hat{H}^{0u}_3\hat{H}^{0d}_+
+b_{33}\hat{H}^{0u}_3\hat{H}^{0d}_3
+h.c.\right],
\label{scalarp1}
\ee
where $b_{++(--)}'=b+b_{++(--)}$,  
$g_{1,2}$ are the gauge coupling constants
for the  $U(1)_{Y}$ and $SU(2)_{L}$ gauge groups,
and $H_\pm$'s are defined in (\ref{Hpm}).
Note that the scalar potential (\ref{scalarp1}) has the same $Z_2$
symmetry as that of  the $b$ sector. 
($H_{+}$'s and $H_{3}$'s are $Z_2$ even, and
$H_{-}$'s are $Z_2$ odd.) Therefore, 
\be
<\hat{H}_{-}^{0u,d}> &=& 0, <\hat{H}_{+}^{0u,d}>
 =\frac{v_{+}^{u,d}}{\sqrt{2}}\exp i \theta_+^{u,d}, 
<\hat{H}_{3}^{0u,d}> 
=\frac{v_3^{u,d}}{\sqrt{2}}\exp i \theta_3^{u,d}
\label{vev1}
\ee
can become a local minimum, where we assume that 
$v_{+}^{u,d}$ and $v_{3}^{u,d}$  are real.
We recall that the $Z_2$ is an accidental symmetry
expect for the $b$ sector.
\footnote{It is accidental in the part of (\ref{scalarp1}) 
coming from the D-terms (the second line).
The $Q_6$ invariant soft scalar mass terms respect automatically
this $Z_2$, although it  is not contained in $Q_6$.
This $Z_2$ is a part of the $O(2)$ symmetry of
the $\mu$ sector, which is only softly broken down to the $Z_2$
in the $b$ sector.} Therefore, the VEV structure (\ref{vev1}) 
is stable against (infinite) renormalization.

We investigate whether the potential energy at the VEV (\ref{vev1}) 
can become negative so that 
$SU(2)_L\times U(1)_Y$ is spontaneously broken.
To this end
we consider the quadratic part of the scalar potential
\be
V^{(2)} &=&
 {\cal H}^I {\cal M}_{IJ} {\cal H}^J~,
\ee
where 
\be
{\cal M} &=&\left(\begin{array}{cccccccc}
m_{H_{+}^{u}}^{2} & 0 & \Re(b_{++}') & -\Im(b_{++}' )
& 0 & 0 & \Re(b_{+3}) & -\Im (b_{+3})\\
0 & m_{H_{+}^{u}}^{2}  & -\Im(b_{++}') & -\Re(b_{++}')
& 0 & 0 & -\Im(b_{+3}) & -\Re (b_{+3})\\
\Re(b_{++}') & -\Im(b_{++}' ) & m_{H_{+}^{d}}^{2} 
& 0& \Re(b_{3+}) & -\Im (b_{3+}) & 0 & 0\\
-\Im(b_{++}') & -\Re(b_{++}' ) &0 & m_{H_{+}^{d}}^{2} 
& -\Im(b_{3+}) & -\Re (b_{3+}) & 0 & 0\\
0 & 0 & \Re(b_{3+}) & -\Im(b_{3+} )& m_{H_{3}^{u}}^{2} 
& 0 & \Re(b_{33}) & -\Im (b_{33})\\
0 & 0 & -\Im(b_{3+}') & -\Re(b_{3+} )
& 0 & m_{H_{3}^{u}}^{2}& -\Im(b_{33}) & -\Re (b_{33})\\
\Re(b_{+3}) & -\Im(b_{+3} ) & 0 & 0&
\Re(b_{33}) & -\Im (b_{33}) & 
m_{H_{3}^{d}}^{2} &0\\
-\Im(b_{+3}) & -\Re(b_{+3} ) & 0 & 0&
-\Im(b_{33}) & -\Re (b_{33}) & 0&
m_{H_{3}^{d}}^{2}
\\\end{array}
\right),\nn\\
 & &
 \label{calM}
\ee
 and
 \be
 {\cal H} &=&(~\Re(\hat{H}_+^{0u}),\Im(\hat{H}_+^{0u}),
 \Re(\hat{H}_+^{0d}),\Im(\hat{H}_+^{0d}),
 \Re(\hat{H}_3^{0u}),\Im(\hat{H}_3^{0u}),
 \Re(\hat{H}_3^{0d}),\Im(\hat{H}_3^{0d})~)~.
 \label{calH}
 \ee
 We find that all the eigenvalues of ${\cal M}$ are doubly generate, 
 and that two orthogonal eigenvectors
 of the same eigenvalue can be always written in the form
 \be
 {\vec u}_A &=& (~u_1, u_2,u_3,u_4,u_5,u_6,u_7,u_8~)
 ~\mbox{and}~
 {\vec u}_B = (~u_2, -u_1,-u_4, u_3, u_6,-u_5,-u_8,u_7~)~.
 \label{us}
 \ee
 This is due to the $U(1)_Y$ gauge invariance:
 All the directions defined by a linear combination
of  $ {\vec u}_A$ 
and $ {\vec u}_B$ are physically equivalent.
If all the imaginary parts of $b$'s vanish, then 
we find $ u_2=u_4=u_6=u_8=0$, which means that
CP can not be spontaneously broken,
because the imaginary parts $\Im ({\cal H}_I)$
along the direction defined by $(~u_1, 0,u_3,0,u_5,0,u_7,0~)$ 
stay at zero. So at least one of $b$'s should be complex so that
CP is  spontaneously broken. \footnote{Spontaneous CP
violation in supersymmetric models and two Higgs
doublet models have been discussed in
 Refs.~\cite{Maekawa:1992un}-\cite{Hugonie:2003yu},
 Ref.~\cite{Maniatis:2007vn} and references therein.}
The product of the four independent eigenvalues is
$\det {\cal M}$. Therefore, if  $\det {\cal M}$ is negative,
one or three  independent eigenvalues are negative.
If $\det {\cal M}$ is positive, there may be zero, 
two or four negative eigenvalues.
In this case one should compute the eigenvalues explicitly.
A local minimum lies along
the direction of a negative eigenvalue.
Further, the potential (\ref{scalarp1}) along
the D-term flat direction should  not be unbounded below. This 
condition requires
\be
 m_{H_{+}^{u}}^{2}+ m_{H_{+}^{d}}^{2}
-2 |b_{++}' | & > & 0~,~
 m_{H_{+}^{u}}^{2}+ m_{H_{+}^{d}}^{2}
-2 |b_{--}' |  >  0~,~
 m_{H_{+}^{u}}^{2}+ m_{H_{3}^{d}}^{2}
-2 |b_{+3} |  >  0~,\nn\\
 m_{H_{3}^{u}}^{2}+ m_{H_{+}^{d}}^{2}
-2 |b_{3+} | & > & 0~,~
 m_{H_{3}^{u}}^{2}+ m_{H_{3}^{d}}^{2}
-2 |b_{33} |  >  0~.
\label{cond1}
\ee

We have to make the flavor changing
neutral Higgs bosons sufficiently heavy to suppress FCNCs.
(This will be discussed in Sec. V.)
So we need a certain fine tuning among the SSB parameters,
because the size of the VEVs is bounded from above.
To achieve this situation, we have to so fine tune the parameters
that one negative eigenvalue at the origin
of the potential becomes very small. \footnote{
By one eigenvalue we mean one of four eigenvalues.
All the eigenvalues are doubly degenerate.}
Then the potential energy falls only slowly
when moving from the origin, and  the quartic terms
in the potential  (\ref{scalarp1})  coming from the D-terms
start to dominate,
so that the energy scale of the VEVs at the bottom of the potential can be much smaller
then the energy scale of the SSB parameters.
Here is such an example:
\be
\Im (b_{++})/\Re (b_{++}') &=&0.747~,~
\Re (b_{33})/\Re (b_{++}') = 0.852~,
~\Im (b_{33})/\Re (b_{++}') = 1.399~,\nn\\
 ~\Re (b_{+3})/\Re (b_{++}') &=&0.667~,
  ~\Im (b_{+3})/\Re (b_{++}')=0.31~, \nn\\
\Re (b_{3+})/\Re (b_{++}') &=&1.3~, ~
\Im (b_{3+})/\Re (b_{++}')=0.42~, \\
 m_{H_{+}^{u}}^{2}/\Re (b_{++}') &=&3.13~,~
  m_{H_{+}^{d}}^{2}/\Re (b_{++}') =2.69~,\nn\\
   m_{H_{3}^{u}}^{2}/\Re (b_{++}') &=&1.39~,~
    m_{H_{3}^{d}}^{2}/\Re (b_{++}') =5.93~.\nn
    \label{example}
\ee
The four independent eigenvalues are
$-5.4\times 10^{-5},~2.27, ~4.16, ~6.70$
in the unit of $b_{++}'$, and
two eigenvectors for the smallest eigenvalue
correspond to
\be
u_1 &= &-0.1070, u_2=0.2232, u_3=0.4091,
u_4=0.3081~,\nn\\
u_5 &=& -0.4216, u_6=0.6636,
u_7=0.2408, u_8=0.0154~,
\label{us-example}
\ee
where $u$'s are defined in (\ref{us}).
Along the direction defined by
(\ref{us-example}) 
 the potential energy falls very slowly when moving from the origin.
So the $SU(2)_L\times U(1)_Y$ invariant point is
a saddle point, and we find that
the size of $\sqrt{b_{++}'}$ may be estimated as
\be
\sqrt{b_{++}'} &\simeq&
\left(\frac{0.13(g^2_2+
3 g_1^2/5)/8}{5.4\times 10^{-5}}\right)^{1/2} 
\times (246~\mbox{GeV})\simeq 3.2 ~\mbox{TeV}~.
\label{finetune}
\ee

CP is also spontaneously broken, because it is not possible
to obtain a vector of the form
$(\bullet, 0, \bullet, 0, \bullet, 0, \bullet, 0)$
through a linear combination of ${\vec u}_A$ and ${\vec u}_B$
for (\ref{us-example}). Therefore, the angle $\theta_q$ 
that enters in the calculation of the CKM 
(given in (\ref{parameters})) is non-zero for (\ref{us-example}).
We find:
\be
\theta_q &=&\theta^u_+-\theta^d_+-\theta^u_3
+\theta^d_3\nn\\
&=&\arctan (u_2/u_1)-\arctan (u_4/u_3)-\arctan (u_6/u_5)
+\arctan (u_8/u_7)\simeq -0.701,
\label{theta-q}
\ee
which is the size of $\theta_q$ we need to produce the correct
CKM parameters as we will see in Sec. V.

\subsection{The heavy neutral Higgs fields}

Now redefine the Higgs fields as follows:
First we define the tilde fields
\be
\tilde{H}^{0u,0d}_{+} &=& \hat{H}^{0u,0d}_{+} \exp- i \theta^{u,d}_{+},~
\tilde{H}^{0u,0d}_{3} = \hat{H}^{0u,0d}_{3} \exp -i \theta^{u,d}_{3},
\ee
and then
\be
\phi^{u}_L &=& \cos\gamma^{u} \tilde{H}_3^{0u}
+\sin\gamma^{u} \tilde{H}_+^{0u},~
\phi^{u}_H = -\sin\gamma^{u} \tilde{H}_3^{0u}
+\cos\gamma^{u} \tilde{H}_+^{0u},
\label{Htilde}
\ee
where
\be
\cos\gamma^u &=& \frac{v_3^u}{\sqrt{(v_3^{u})^2+(v_+^{u})^2}},~
\sin\gamma^u = \frac{v_+^u}{\sqrt{(v_3^{u})^2+(v_+^{u})^2}},
\label{cosgamma}
\ee
and similarly for the down sector. As we see from (\ref{cosgamma}),
only $\phi^{u}_L$ and $\phi^{d}_L$ have a nonvanishing VEV, which we denote
by
\be
<\phi^{u,d}_L>
&=& \frac{\sqrt{(v^{u,d}_3)^2+(v^{u,d}_+)^2}}{\sqrt{2}}=
\frac{v_{u,d}}{\sqrt{2}}.
\ee
The neutral light and heavy scalars of the MSSM are given by
\be
\frac{1}{\sqrt{2}}( v + h ) &=&\mbox{Re}( \phi^{d*}_L)\cos\beta
+\mbox{Re} (\phi^{u}_L)\sin\beta,\\
\frac{1}{\sqrt{2}} (H+i A) &=&- (\phi^{d*}_L)\sin\beta
+ (\phi^{u}_L)\cos\beta,
\ee
where as in the MSSM
\be
v &=&\sqrt{v_u^2+v_d^2}, ~\tan\beta=\frac{v_u}{v_d}~.
\label{beta}
\ee

As in the case of the MSSM, the  couplings of $\phi^{u,d}_L$ are
flavor-diagonal, and so we do not have to  consider them below when
discussing FCNCs.
Therefore, only the heavy fields
$\hat{H}^{0u,0d}_{-}=\phi_-^{u,d}$ and $\phi^{u,d}_H$ 
can have flavor-changing couplings. Their mass matrix  can be written as
\be
\left(
\begin{array}{cccc}
m^2_{\phi_I^u} & 0 & 0& b_I^*\\
0 & m^2_{\phi_I^u}  & b_I&0 \\
0 & b_I^* & m^2_{\phi_I^d} & 0\\
b_I & 0 & 0& m^2_{\phi_I^d}\\
\end{array}\right)~
\label{massphi}
\ee
in the
$(\phi_I^{u}, \phi_I^{u*}, \phi_I^{d}, \phi_I^{d*})$  basis,
where $I=-,H$,
\be
m^2_{\phi_-^{u,d}} &=& m^2_{H_-^{u,d}}, b_-=b_{--}'~,
~m^2_{\phi_H^{u,d}} =
m^2_{H_+^{u,d}}\cos^2\gamma^{u,d}+
m^2_{H_3^{u,d}}\sin^2\gamma^{u,d}~,\nn\\
b_H &=&b_{++}' e^{-i(\theta_+^u+\theta_+^d)} 
\cos\gamma^u \cos\gamma^d-
b_{+3} \cos\gamma^u \sin\gamma^de^{-i(\theta_+^u+\theta_3^d)} 
\label{mandb}\\
& -&
b_{3+} \sin\gamma^u \cos\gamma^de^{-i(\theta_3^u+\theta_+^d)} +
b_{33} \sin\gamma^u \sin\gamma^d e^{-i(\theta_3^u+\theta_3^d)} ~,\nn
\ee
and the mass parameters on the rhs are given in (\ref{scalarp1})
and $\gamma^{u,d}$ are defined in (\ref{cosgamma}).
The inverse of the matrix (\ref{massphi}) is given by
\be
\frac{1}{(M_{I1})^2( M_{I2})^2}
\left(\begin{array}{cccc}
m^2_{\phi_I^d} & 0 & 0& -b_I^*\\
0 & m^2_{\phi_I^d}  & -b_I&0 \\
0 & -b_I^* & m^2_{\phi_I^u} & 0\\
-b_I & 0 & 0& m^2_{\phi_I^u}
\end{array}\right)~~,~~(I=-,H)~,
\label{inverse}
\ee
where $M_{1,2}$ are approximate pole masses and given by
\be
(M_{I1(2)})^2 &=&\frac{1}{2}(m^2_{\phi_I^u}+m^2_{\phi_I^d})
\left(1+(-)\left[\frac{4 |b_I|^2
+(m^2_{\phi_I^u}-m^2_{\phi_I^d})^2}{
(m^2_{\phi_I^u}+m^2_{\phi_I^d})^2}\right]^{1/2}\right)~,
\label{M1M2}
\ee
and we find
\be
(M_{I1})^2 (M_{I2})^2 &=&
-|b_I|^2+m^2_{\phi_I^u}m^2_{\phi_I^d}.
\label{m1m2}
\ee
(\ref{inverse}) is the inverse propagator at the zero momentum.
We will be using it later on.
For the parameter values  in the example (\ref{us-example}) we find
\be
\tan \gamma^u &=& 0.315 ~,~\tan \gamma^d =2.122 ~,~
\tan \beta = -1.456~,\nn\\
M_{H1} &=& 2.31 \sqrt{b_{++}'}
\simeq 7.3~~\mbox{TeV} ~,~
M_{H2}=1.72 \sqrt{b_{++}'}\simeq 5.5~~\mbox{TeV} ~,
\label{M12}
\ee
where we have used (\ref{finetune}).
 So, what we have numerically shown in A and B in this section is that
 it is possible to fine tune the SSB parameters
 so as to make the heavy Higgs bosons much heavier than
$M_Z$ (see (\ref{M12})) and at the same time to obtain
a desired size of spontaneous CP violation (see (\ref{theta-q})).

\section{FCNCs}

\subsection{The physical  quarks and leptons}
From the Yukawa interactions (\ref{wQ})
and (\ref{wL})
along with the  form of the VEVs (\ref{vev1})
we obtain the fermion mass matrices.
 
 \subsubsection{Quark sector}
 The quark mass matrices are given by
\be
{\bf m}^{u} &=&\frac{1}{2}\left(\begin{array}{ccc}
0 &\sqrt{2} Y_c^{u} v_3^u e^{-i\theta^u_3}
& Y_b^{u}  v_+^u e^{-i \theta_+^u}\\
\sqrt{2} Y_c^{u}v_3^u e^{-i \theta^u_3} & 0 &
 Y_b^{u} v_+^ue^{-i  \theta_+^u} \\
-Y_{b'}^{u} v_+^ue^{-i \theta_+^u} & 
Y_{b'}^{u}v_+^u e^{-i \theta_+^u} & 
\sqrt{2} Y_a^{u}  v_3^u e^{-i \theta^u_3} \\
\end{array}\right),
\label{mu}\\
{\bf m}^{d} &=&\frac{1}{2}\left(\begin{array}{ccc}
0 & \sqrt{2}Y_c^{d} v_3^d e^{-i \theta^d_3}
&Y_b^{d}  v_+^d e^{-i \theta_+^d}\\
\sqrt{2} Y_c^{d}v_3^d e^{-i \theta^d_3} & 0 &
Y_b^{d} v_+^de^{-i \theta_+^d} \\
-Y_{b'}^{d} v_+^de^{-i \theta_+^d} & 
 Y_{b'}^{d}v_+^d e^{-i \theta_+^d} & 
\sqrt{2}Y_a^{d}  v_3^d e^{-i \theta^d_3} \\
\end{array}\right).
\label{md}
\ee
Then using the phase matrices defined below
\be
R_L &=&  \frac{1}{\sqrt{2}}\left( \begin{array}{ccc}
1 & 1 & 0\\-1 & 1& 0\\
 0 & 0 &\sqrt{2}
\end{array}\right),~
R_R =  \frac{1}{\sqrt{2}}\left( \begin{array}{ccc}
-1 & -1 & 0\\-1 & 1& 0\\
 0 & 0 &\sqrt{2}
\end{array}\right),
\label{R}\\
P_L^u &=&    \left( \begin{array}{ccc}
1 & 0 & 0\\0 & \exp (i2\Delta \theta^{u}) & 0\\
 0 & 0&\exp (i\Delta \theta^{u})
\end{array}\right),
\label{Pu}\\
P_R^u &=&
\left( \begin{array}{ccc}
\exp (i2\Delta \theta^{u})  & 0 & 0\\0 & 1& 0\\
 0 & 0&\exp (i\Delta \theta^{u})
\end{array}\right)  
\exp ( i \theta_3^u),
\label{Puc}\\
\Delta \theta^{u} & = &
 \theta_3^u-\theta^u_+,
 \label{dtheta}
\ee
and similarly for the down sector,
 we can bring ${\bf m}^u$ into a real form
\be
{\bf \hat{m}}^u &=& P_L^{u \dag}R^T_L {\bf m}^u R_R P_R^u 
=m_t\left(\begin{array}{ccc}
0 & q_u/y_u & 0  \\ -q_u/y_u & 0 & b_u\\
0 &   b_u'  &  y^2_u \\
\end{array}\right)~.
\label{muhat}
\ee
The mass matrix ${\bf \hat{m}}^u$ can then be diagonalized as
\be
O^{uT}_L {\bf \hat{m}}^u O_R^u &=&
\left( \begin{array}{ccc}m_u & 0 & \\ 0 & m_c & 0\\ 0 & 0 & m_t
\end{array}\right)~, 
\label{omo}
\ee
and similarly for ${\bf m}^d$, where
$ O_{L,R}^{u,d}$ are orthogonal matrices.
So the mass eigenstates $u_{iL}'=(u_{L}', c_{L}',t_{L}')$ 
etc. can be  obtained from
\be
u_L &=& U_L^u u_L'~,u_R = U_R^u u_R',
d_L = U_L^d d_L',~d_R = U_R^d d_R',
\ee
where
\be
U_{uL(R)} &=&  R_{L(R)} P_{L(R)}^u O_{L(R)}^u~.
\ee
Therefore, the CKM matrix $V_{\rm CKM}$ is
 given by 
 \be
 V_{\rm CKM} &=& O^{uT}_LR_L^T R_R P_L^{u\dag}  O_L^d=
 O^{uT}_L P_q O_L^d~
 \label{vckm}
 \ee
 where
 \be
P_q &=&   P_L^{u\dag} 
 P_L^{d} =
  \left( \begin{array}{ccc}
1 & 0 & 0\\0 & \exp (i2\Delta \theta_q) & 0\\
 0 & 0&\exp (i\Delta \theta_q)
\end{array}\right)~.
\label{Pq}
 \ee
For the set of the parameters
 \be
\theta_q & = &\theta^d_3-\theta^d_+-\theta^u_3
+\theta^u_+  =-0.7,
q_u=0.0001799,
b_u=0.05979,
b'_u=0.07054,\nn\\
y_u &=&0.99786,
q_d =0.003784,
b_d=0.03268,
b'_d=0.4620,
y_d=-0.9415,
\label{parameters}
\ee
we obtain
\be
m_u/m_t &=& 0.766\times 10^{-5},
m_c/m_t=4.23 \times 10^{-3},
m_d/m_b=0.895 \times 10^{-3},
m_s/m_b=1.60 \times 10^{-2},
\nn\\
|V_{\rm CKM}| &=&
\left( \begin{array}{ccc}
0.9740 & 0.2266  & 0.00362
\\  0.2265   & 0.9731& 0.0417
  \\ 0.00849 &0.0410& 0.9991
\end{array}\right),~
|V_{td}/V_{ts}| =0.207~,
\label{predQ}\\
\sin 2\beta (\phi_1) &=&0.690~,~\gamma(\phi_3)=63.4^o~.
\label{ckm-parameters}
\ee
The experimental values to be compared are 
\cite{Yao:2006px} (see also \cite{Bona:2006ah}):
\be
|V_{\rm CKM}^{\rm exp}| &=&
\left( \begin{array}{ccc}
0.97383
\begin{array}{c}\vspace{-2mm}+0.00024\\
-0.00023\end{array} & 0.2272 \begin{array}{c}\vspace{-2mm}+0.0010\\
-0.0010\end{array}
 & 0.00396 \begin{array}{c}\vspace{-2mm}+0.00009\\
-0.00009\end{array}
\\  0.2271\begin{array}{c}\vspace{-2mm}+0.0010\\
-0.0010\end{array}
 & 0.97296
\begin{array}{c}\vspace{-2mm}+0.00024\\
-0.00024\end{array}   & 0.04221\begin{array}{c}\vspace{-2mm}+0.00010\\
-0.00080\end{array}
  \\ 0.00814\begin{array}{c}\vspace{-2mm}+0.00032\\
-0.00064\end{array}
  &0.04161 
\begin{array}{c}\vspace{-2mm}+0.00012\\
-0.00078\end{array} & 0.999100 \begin{array}{c}\vspace{-2mm}+0.000034\\
-0.000004\end{array}\end{array}\right),\nn\\
\sin 2\beta (\phi_1)&=&0.687\pm0.032,~,~
|V_{td}/V_{ts}| =0.208\begin{array}{c}\vspace{-2mm}+0.008\\
-0.006\end{array}~.
 \ee
 The quark masses at $M_Z$ are given by \cite{Kim:2004ki}
\be
m_u/m_d &=& 0.541\pm 0.086~(0.51)~,
~m_s/m_d = 18.9\pm 1.6~(17.9),\nn\\
~m_c  &=&0.73\pm 0.17~(0.74)~\mbox{GeV}~,
~m_s =0.058\pm 0.015~(0.046)~\mbox{GeV},\nn\\
~m_t &=& 175 \pm 6 ~\mbox{GeV}~,
~m_b = 2.91\pm 0.07~\mbox{GeV},
\label{qmass}
\ee
where the values in the parentheses are the theoretical values obtained
from  (\ref{predQ}) 
for $m_t= 174  $ GeV and $m_b=2.9  $ GeV. So, we see that the model can well reproduce the experimentally measured 
parameters.

The orthogonal matrices (\ref{omo}) are found to be
\be
 O_{uL}   &\simeq & \left(
\begin{array}{ccc}
0.9991 & -0.04252 & 1.269\times 10^{-5} \\
0.04244 & 0.9973 & 0.05964\\
-2.548\times 10^{-3} &  -0.05958& 0.9982
\end{array}\right),
\label{UuL}
\\
O_{uR} & \simeq   &\left(
\begin{array}{ccc}
-0.9991 &  -0.04255 & -1.075\times 10^{-5}\\
0.04244 & -0.9966 & 0.07042 \\
-3.007\times 10^{-3} &0.07035 & 0.9975
\end{array}\right),
\label{UuR}
\ee
\be
O_{dL}  &\simeq& \left(
\begin{array}{ccc}
0.9764 & 0.2160 &  -1.856\times 10^{-3}\\
-0.2159 & 0.9760 & 0.02899\\
8.074\times 10^{-3} &-0.02790 &  0.9996
\end{array}\right),
\label{UdL} \\
O_{dR}  &\simeq& \left(
\begin{array}{ccc}
-0.9695 &  0.2452 &  1.165\times 10^{-4}\\
-0.2174 &-0.8599 &  0.4618\\
0.1133 & 0.4477 & 0.8870
\end{array}\right)~.
\label{UdR}
\ee

\subsubsection{Lepton sector}
The charged lepton  mass matrix
becomes
\be
{\bf m}_{e} = \left( \begin{array}{ccc}
-m_{2} & m_{2} & m_{5} 
\\  m_{2} & m_{2} &m_{5}
  \\ m_{4} & m_{4}&  0
\end{array}\right)\exp (-i \theta_+^d),
\label{mlepton}
\ee
where
\be
m_{2} &=& \frac{1}{2} Y_c^e v_+^d, ~
m_{4}=\frac{1}{2} Y_{b'}^e v_+^d, ~
m_{5}=\frac{1}{2} Y_b^e v_+^d.
\ee
The phase $\exp (-i \theta_+^d)$
 can be  rotated away, and  all the mass parameters
appearing in (\ref{mlepton}) are real.
Diagonalization of the mass matrices is straightforward.

We would like to mention that 
the model has many predictions in this sector, because
there are only four parameters to describe
three light neutrino masses, three angles and three CP violating phases
of $V_{MNS}$. Since the details of the predictions are
presented in 
 Refs.~\cite{Kubo:2003iw,Kubo:2003pd,Felix:2006pn}, 
 we do not repeat them here again.
\footnote{See also \cite{Kajiyama:2007pr} for the predictions
of the model on R parity violating processes. 
The leptonic sector of the present model is
basically the same as the model of 
\cite{Kubo:2003iw,Kubo:2003pd},
except for the spontaneous breaking of CP, which
reduces one more independent phase in the leptonic sector.}
Furthermore, the FCNC processes in the lepton sector
have been very recently analyzed in details in  Ref.~\cite{Mondragon:2007af}, concluding that the model predictions of
tree-level FCNC processes are  at least
five orders of magnitude smaller than the experimental upper bounds
(The mass of the heavy neutral Higgs fields are assumed to be $120$ GeV.)
For instance, the branching fraction for 
$\mu\to e \gamma$  is
seven orders of magnitude smaller than the expected experimental 
sensitivity \cite{Mondragon:2007af}. Therefore, we shall not consider 
FCNCs in the leptonic sector in the following discussions.

\subsection{CP violations and FCNCs
in the SSB sector}

If three generations of a family 
have the one+two structure, then 
the soft scalar mass matrices for the sfermions have a diagonal form
(\ref{m2}):
\be
{\bf \tilde{m}^2}_{aLL(RR)} =
{m}^2_{\tilde{a}} \left(
\begin{array}{ccc}
a_{L(R)}^{a} & 0 & 0 \\
0 & a_{L(R)}^{a} & 0 \\
0 & 0 & b_{L(R)}^{a}
\end{array}
\right)~~(a=u,d,e)~,
\label{scalarmass}
\ee
where ${m}_{\tilde{a}}$ denote the average of the  squark 
and slepton masses, respectively,  and $(a_{L(R)}, b_{L(R)})$ are
dimensionless free real parameters of $O(1)$.
Because of the $Q_6$ flavor symmetry in 
the trilinear  interactions,
all the soft left-right mass matrices assume the form
\be
\left({\bf \tilde{m}^2}_{aLR}\right)_{ij} 
&=&
A_{ij}^a\left( {\bf m}^a \right)_{ij} 
~~(a=u,d,e)~,
\label{Aterm}
\ee
where $A_{ij}^{a}$ are free parameters of dimension one (see  (\ref{hijk})).
They are also real, because we impose CP invariance in the tri-linear couplings.

The quantities \cite{Hall:1985dx,Gabbiani:1988rb}
\be
\Delta_{LL(RR)}^{a} &=&
U_{aL}^{\dagger} ~{\bf \tilde{m}^2}_{aLL(RR)}~
 U_{aL(R)}~\mbox{and}~
\Delta_{LR}^{a} =
U_{aL}^{\dagger}~ 
{\bf \tilde{m}^2}_{aLR} ~U_{aR}
\label{Delta1}
\ee
in the super CKM basis  are used widely to parameterize FCNCs and CP violations coming
from the SSB sector, where the unitary matrices 
$ U$'s are given in (\ref{UuL})-(\ref{UdR}).

\subsubsection{CP violations}
The imaginary parts of
$\Delta$'s  (\ref{Delta1})
contribute to CP violating processes in the SSB sector.
Recall that the soft scalar mass matrices ${\bf m}^2_{aLL,RR}$ are real,
because they are diagonal, and that the phases of
${\bf m}^2_{aLR}$ come from the complex VEVs (\ref{vev1}),
because CP is only spontaneously broken in this sector.
The unitary matrices $U$'s are complex, and so $\Delta$'s can be
complex, too. Note that the unitary matrices have the form
$U=R P O$, where only $P$'s  (given in (\ref{Pu})) are
complex. Since $P$'s are diagonal, they commute with
${\bf m}^2_{aLL,RR}$, so that $\Delta_{LL,RR}^a$ have no
imaginary part.  Further ${\bf m^2}_{aLR}$ has
the same phase structure as the corresponding fermion mass matrix
$ {\bf m}^a$, which can be made real according to (\ref{R})-
(\ref{muhat}).
Therefore, $\Delta_{LR}^a$, too, are real. Consequently,
there is no CP violation originating from the SSB sector.
The stringent constraints on $\Delta$'s (\ref{Delta1})
 coming from
the EDMs \cite{Abel:2001vy,Endo:2003te,Hisano:2004tf} 
are automatically satisfied in this way
of phase alignment.
\footnote{This does not mean that there is no CP violation
in the SSB sector.  Due to the existence of the multiple Higgs
fields, there are one-loop diagrams contributing
to the EDMs, even if all the SSB parameters are real.
The diagrams typically contain the $b$ terms, and 
we find that in the case of the present model
$b_-  << m_{H_-^{u,d}}^2~,~
b_H < m_{\phi_H^{u,d}}^2$
(given  in (\ref{mandb} )) should be satisfied
to satisfy the experimental constraints.}

\subsubsection{FCNC}
In  Refs.~\cite{Hall:1985dx,Gabbiani:1988rb},
\cite{Gabbiani:1996hi}-\cite{Golowich:2007ka}
\cite{Abel:2001vy}-\cite{Hisano:2004tf}, experimental bounds  on the dimensionless quantities
\be
\delta^{a}_{LL,RR,LR} &=& 
\Delta^{a}_{LL,RR,LR}/{m}^2_{\tilde a}~~~(a=u,d),
\ee
are given.
The theoretical values of $\delta$'s for the present model 
have been calculated in  Ref.~\cite{Kajiyama:2005rk} as a function of 
the average sfermion masses and fine tuning parameters.
The results may be summarized as follows.
For the slepton sector where the average slepton
mass $m_{\tilde e}$ is assumed to be $500$ GeV, the theoretical 
values of $(\delta^{\ell}_{ij})_{LL,RR,LR}$,
except for $(\delta^{\ell}_{12})_{LL}$,  are several
orders of magnitude smaller than the current experimental
bounds, while 
$(\delta^{\ell}_{12})_{LL}$ is of the same order
as that of the experimental bound which comes from
$\mu\to e \gamma$.
In the squark sector, we find:

\vskip 0.5cm
\underline{\bf Up quark  sector:}
\be
(\delta^u_{12})_{LL}
&=&  (\delta^u_{21})_{LL}\simeq 
-1.5 \times 10^{-4}~\Delta a_L^{q}, \nn\\
(\delta^u_{12})_{RR}
&=& (\delta^u_{21})_{RR}
\simeq -2.1 \times 10^{-4} ~\Delta a_R^{u}, \label{deltau}~,\\
(\delta^{u}_{12})_{LR}
&\simeq&-(\delta^{u}_{21})_{LR}\simeq 
6.2 \times 10^{-5} \left(-\tilde{A}_{a}^{u}+\tilde{A}_{b}^{u}+
\tilde{A}_{b'}^{u}-\tilde{A}_{c}^{u}
\right)~
\left( \frac{500 ~\mbox{GeV}}{m_{\tilde{q}} }\right) ~,\nn
\ee

\vskip 0.5cm
\underline{\bf Down quark  sector}
\be
(\delta^d_{12})_{LL}
&=&  (\delta^d_{21})_{LL}\simeq 
2.2 \times 10^{-4}~\Delta a_L^{q}, \nn\\
(\delta^d_{13})_{LL}
&=& (\delta^d_{31})_{LL}
\simeq  -8.1\times 10^{-3}~\Delta a_L^q, \nn\\
(\delta^d_{23})_{LL}
&=& (\delta^d_{32})_{LL}
\simeq  2.8 \times 10^{-2}~\Delta a_L^{q},\nn\\
(\delta^d_{12})_{RR}
&=& (\delta^d_{21})_{RR}
\simeq -5.1 \times 10^{-2} ~\Delta a_R^{d}, \label{deltad}
\\
(\delta^d_{13})_{RR}
&=& (\delta^d_{31})_{RR}
\simeq -0.1~\Delta a_R^{d}, \nn\\
(\delta^d_{23})_{RR}
&=& (\delta^d_{32})_{RR}
\simeq -0.4~\Delta a_R^{d},\nn
\ee
where 
\be
\Delta a_{L}^{q} &=&a_{L}^{q}-b_{L}^{q},~
\Delta a_{R}^{a} =a_{R}^{a}-b_{R}^{a},~
\tilde{A}_{i}^{a} = \frac{A_{i}^{a}}{{m}_{\tilde q}}~~(a=u,d).
\label{deltaAt1}
 \ee
These parameters, $a_{L,R}$ and $\tilde{A}_i$, are free dimensionless 
parameters, so that they are $O(1)$ if we do not 
fine tune them.
The most stringent constraint in the up-sector comes from
$\Delta M_D$ \cite{Ciuchini:2007cw,Golowich:2007ka}:
\be
\Delta M_D &=>& |(\delta^u_{12})_{LL}|,|(\delta^u_{12})_{RR}|
\lsim 6 \times 10^{-2}~,~
|(\delta^u_{12})_{LR}|,|(\delta^u_{21})_{LR}|
\lsim  10^{-2}
\ee
for $m_{\tilde q} =0.5 $ TeV.
As we can see from (\ref{deltau}) this constraint
can be satisfied without a fine-tuning.
As for the down-sector we have to satisfy the
constraints coming from $\Delta M_K,
\Delta M_{B_s}$ and $\Delta M_{B_d}$
   \cite{Gabbiani:1996hi, Ciuchini:2006dx}:
\be
\Delta M_K &=>&
 |(\delta^d_{12})_{LL}|~,~|(\delta^d_{12})_{RR}|~,~
 |(\delta^d_{12})_{LR}|~,~|(\delta^d_{21})_{LR}|
\lsim  10^{-3}\\
\Delta M_{B_d} &=>&
 |(\delta^d_{13})_{LL}|~,~|(\delta^d_{13})_{RR}|~,~
 |(\delta^d_{13})_{LR}|~,~|(\delta^d_{31})_{LR}|
\lsim  10^{-2}\\
\Delta M_{B_s} &=>&
 |(\delta^d_{23})_{LL}|~,~|(\delta^d_{23})_{RR}|~,~
 |(\delta^d_{23})_{LR}|~,~|(\delta^d_{32})_{LR}|
\lsim  10^{-1}
\ee
Comparing these constraints with (\ref{deltad}) we see
that $\Delta a_R^d$ should be fine tuned at the level of few \%.
\footnote{We find that, as in the case of
$(\delta^u_{12})_{LR}$ of (\ref{deltau}), the left-right insertions $|(\delta^d_{12,21,13,31,23,32})_{LR}|$
are much smaller than these constraints.}
In the next subsections we assume that $\Delta a_R^d$ is so small
that only the heavy flavor-changing-neutral Higgs fields
contribute to the mass differences of the neutral mesons.

\subsection{Flavor changing neutral Higgs couplings}

In Sec. IV we found that only the Higgs fields 
$\phi_{H,-}^{u,d}$ have flavor changing neutral couplings to
the fermions, and that they have a definite form of mixing
(see (\ref{massphi})).
These are consequences of the $Z_2$ symmetry which is
a part of the  $O(2)$ flavor symmetry in the $\mu$ sector
(as discussed in Sec. III. B).
 In the basis of the fermion 
mass eigenstates these Higgs couplings have the following form:
\be
{\cal L}_{FCNC}
&=& -\left[~Y_{ij}^{uH}\phi_H^{u}
+Y_{ij}^{u-}\phi^{u}_-~\right]^*
\overline{u}_{iL}' u_{jR}'
-\left[~Y_{ij}^{dH}\phi_H^{d}+Y_{ij}^{d-}
\phi^{d}_-~\right]^*
\overline{d}_{iL}' d_{jR}'\nn\\
& & -\left[Y_{ij}^{eH}\phi_H^{d}+Y_{ij}^{e-}
\phi^{d}_-\right]^* \overline{e}_{iL}' e_{jR}'+h.c.~,
\label{Lfcnc}
\ee
where the Higgs fields are defined in (\ref{Htilde}), and
\be
{ \bf Y}^{uH} &=&U^{u\dag}_L\left[\frac{1}{\sqrt{2}}\cos\gamma^u
 e^{-i \theta^u_+}( {\bf Y}^{u1}+{\bf Y}^{u2})
-\sin\gamma^u e^{-i \theta_3^u} {\bf Y}^{u3}\right] U^{u}_R,\nn\\
&=& O^{u\dag}_L\left[\frac{1}{\sqrt{2}}\cos\gamma^u
( {\bf Y}^{u1}+{\bf Y}^{u2})
-\sin\gamma^u  {\bf Y}^{u3}\right] O^{u}_R,\\
{ \bf Y}^{dH} &=&U^{d\dag}_L\left[\frac{1}{\sqrt{2}}\cos\gamma^d
 e^{-i \theta^d_+}( {\bf Y}^{d1}+{\bf Y}^{d2})
-\sin\gamma^d e^{-i \theta_3^d} {\bf Y}^{d3}\right] U^{d}_R,\nn\\
&=& O^{d\dag}_L\left[\frac{1}{\sqrt{2}}\cos\gamma^d
( {\bf Y}^{d1}+{\bf Y}^{d2})
-\sin\gamma^d {\bf Y}^{d3}\right] O^{d}_R,\\
{ \bf Y}^{I-} &=&U^{I\dag}_L\left[\frac{1}{\sqrt{2}}
( {\bf Y}^{I1}-{\bf Y}^{I2})\right] U^{I}_R ~~
(I=u,d)~.
\ee
The Yukawa matrices ${\bf Y}^{u1}$ etc. are given in (\ref{Yuq}), and 
the  unitary matrices are given in (\ref{R})-
(\ref{muhat}) and (\ref{UuL})-(\ref{UdR}).

The present model is consistent with the experimental observations
in a certain region in the parameter space of the Yukawa
couplings. An example of the
choice of the nine parameters is given in (\ref{parameters}),
where we emphasize that this set of the
nine parameters describe 10 physical independent quantities
of the SM; six quark masses and four CKM parameters.
Therefore, the consistent region in the  space of the Yukawa
couplings is very restricted, and
we will be using only this set of the 
parameter values in the following discussion.
Accordingly, for the values given in (\ref{parameters})
 we find the actual size of the Yukawa
couplings:
\be
Y_a^u &=& \frac{\sqrt{2}m_t y_u ^2}{v_u\cos\gamma^u}
\simeq \frac{0.9957}{\sin\beta\cos\gamma^u}~,~
Y_b^u = \frac{\sqrt{2}m_t b_u}{v_u\sin\gamma^u}
\simeq \frac{0.05979}{\sin\beta\sin\gamma^u}~,\\
Y_{b'}^u &=& \frac{\sqrt{2}m_t b'_u}{v_u\sin\gamma^u}\simeq
\frac{0.07054}{\sin\beta\sin\gamma^u}~,~
Y_c^u = \frac{\sqrt{2}m_t q_u}{y_u v_u\cos\gamma^u}\simeq
\frac{1.802\times 10^{-4}}{\sin\beta\cos\gamma^u}~,\\
Y_a^d &=& \frac{\sqrt{2}m_b y_d ^2}{v_d\cos\gamma^d}\simeq
\frac{0.01478}{\cos\beta\cos\gamma^d}~,~
Y_b^d = \frac{\sqrt{2}m_b b_d}{v_d\sin\gamma^d}\simeq
\frac{5.449\times 10^{-4}}{\cos\beta\sin\gamma^d}~,\\
Y_{b'}^d &=& \frac{\sqrt{2}m_b  b'_d}{v_d\sin\gamma^d}\simeq
\frac{7.702\times 10^{-3}}{\cos\beta\sin\gamma^d}~,~
Y_c^d = \frac{\sqrt{2}m_b q_d}{y_d v_d\cos\gamma^d}\simeq
\frac{-6.701\times 10^{-5}}{\cos\beta\cos\gamma^d}~,
\ee
where  $\gamma$'s and 
$\beta$ are given in (\ref{cosgamma}) and (\ref{beta}), respectively,
and we have used: $m_t=174$ GeV, $m_b=2.9$ GeV and
$v=\sqrt{v_u^2+v_d^2}=246$ GeV.
These parameters are defined in the  $\overline{\rm MS}$
  scheme and evaluated at the scale $M_Z$.
With these numerical values we then obtain:
\be
{\bf Y}^{uH}
&\simeq&\frac{1}{\tan\gamma^u\sin\beta}\left( \begin{array}{ccc}
-2.65\times 10^{-4} &  
3.22\times 10^{-3} &0.0439
\\-3.22\times 10^{-3} &5.68\times 10^{-3}
&0.0400  \\0.0519
 &-0.0473  &6.02\times 10^{-3}
\end{array}\right)\nn\\
& &-\frac{\tan\gamma^u}{\sin\beta}\left( \begin{array}{ccc}
7.63\times 10^{-6} &  
-3.58\times 10^{-4} &-2.52\times 10^{-3}
\\-1.54\times 10^{-6}
 &-4.17\times 10^{-3}
&-0.0592
  \\-2.99\times 10^{-3}
 &0.0699  &0.991
\end{array}\right)~,
\label{YuH}
\ee
\be
{\bf Y}^{u-}
&\simeq&\frac{\exp i(2\theta^u_3-\theta^u_+)}
{\sin\gamma^u\sin\beta}\left( \begin{array}{ccc}
0 &   -4.21\times 10^{-3}
& -0.0596
\\-4.21\times 10^{-3} & 0 &2.54\times 10^{-3}
  \\0.0704
   &3.00\times 10^{-3}
   &  0   \end{array}\right)~,
   \label{Yum}
   \ee
\be
{\bf Y}^{dH}
&\simeq&\frac{1}{\tan\gamma^d\cos\beta}\left( \begin{array}{ccc}
6.63\times 10^{-5} &  
8.26\times 10^{-5} &2.80\times 10^{-4}
\\-6.224\times 10^{-5}
 &3.74\times 10^{-4}
&3.37\times 10^{-4}
  \\4.10\times 10^{-3}
 &-6.01\times 10^{-3}  &2.52\times 10^{-3}
\end{array}\right)\nn\\
& &-\frac{\tan\gamma^d}{\cos\beta}\left( \begin{array}{ccc}
1.37\times 10^{-5} &  
1.13\times 10^{-4} &7.56\times 10^{-5}
\\1.98\times 10^{-5}
 &-1.88\times 10^{-4}
&-3.72\times 10^{-4}
  \\1.67\times 10^{-3}
 &6.61\times 10^{-3}  &0.0131
\end{array}\right)~,
\label{YdH}
\ee
\be
{\bf Y}^{d-}
&\simeq&\frac{\exp i(2\theta^u_3-\theta^u_+)}
{\sin\gamma^d\cos\beta}\left( \begin{array}{ccc}
0 &  
-2.53\times 10^{-4}
& -4.72\times 10^{-4}
\\-2.22\times 10^{-4}
 & 0 &-1.04\times 10^{-4}
  \\7.46\times 10^{-3}
   &-1.89\times 10^{-3}
   &  0   \end{array}\right)~.
   \label{Ydm}
   \ee
The phases appearing in the matrices are given in (\ref{mu})
and (\ref{md}).
As we can see from these Yukawa matrices the size of the entries
is fixed once the ratios of the VEVs ($\sin\beta, \sin\gamma^u$ etc.)
are fixed. For the down-type Yukawa matrices (\ref{YdH})
and (\ref{Ydm}), for instance,
all the entries
(except the $(3,3)$ entry) are at most  $O(10^{-3})$.
All these facts originate from the flavor symmetries
of the model. Needless to say that in multi-Higgs models without
 a flavor symmetry this situation is completely different.

\subsection{FCNC}
The most severe FCNC constraints on the theory
come from the mass differences in the neutral meson systems;
$\Delta M_D, \Delta M_K,
\Delta M_{B_s}$ and 
$\Delta M_{B_d}$. \footnote{The contribution to 
$\epsilon'/\epsilon$ is negligibly small,
at most $O([10^{-7}/\alpha_S^2][ \tilde{m}_q^2/M^2])$,
where $\sim 10^{-7}$
originates from the Yukawa couplings relevant to this
quantity, and $\tilde{m}_q$ and $M$ stand for
the generic average squark and  charged Higgs masses.
See \cite{Grimus:2007if} and references therein
for the constraint  from the oblique corrections 
due to multiple $SU(2)_L$ doublet Higgs fields.}
The Yukawa  interaction terms  that contribute to
them can be found from (\ref{Lfcnc}):
\be
{\cal L}_{\Delta M_B} &=&
-\left[ Y_{uc}^{uH}\phi^{u}_H+Y_{uc}^{u-}
\phi^{u}_-  \right]^*
\overline{u}_{L} c_{R}
-\left[ Y_{cu}^{uH}\phi^{u}_H+Y_{cu}^{u-}
\phi^{u}_-  \right]\overline{u}_{R} c_{L}\nn\\
&-&\left[ Y_{sd}^{dH}\phi^{d}_H+Y_{sd}^{d-}
\phi^{d}_-  \right]^*
\overline{s}_{L} d_{R}
-\left[ Y_{ds}^{dH}\phi^{d}_H+Y_{ds}^{d-}
\phi^{d}_-  \right]\overline{s}_{R} b_{L}\nn\\
&-&\left[ Y_{bd}^{dH}\phi^{d}_H+Y_{bd}^{d-}
\phi^{d}_-  \right]^*
\overline{b}_{L}d_{R}
-\left[ Y_{db}^{dH}\phi^{d}_H+Y_{db}^{d-}
\phi^{d}_-  \right]\overline{b}_{R} d_{L}\nn\\
&-&\left[ Y_{bs}^{dH}\phi^{d}_H+Y_{bs}^{d-}
\phi^{d}_-  \right]^*\overline{b}_{L}s_{R}
-\left[ Y_{sb}^{dH}\phi^{d}_H+Y_{sb}^{d-}
\phi^{d}_-  \right]\overline{b}_{R}s_{L}~,
\label{LDM}
\ee
where the values of the Yukawa couplings 
can be read off from (\ref{YuH})-(\ref{Ydm}).
In (\ref{LDM}) we have dropped the prime on the fields,
which was indicating the mass eigenstate.
As we can see from (\ref{inverse}),
 no $\phi-\phi$ and $\phi^*-\phi^*$ type propagators
 contribute to the mass differences.
 So, only 
 the $\phi-\phi^*$ type propagators
 can contribute, implying the  phases in the Yukawa couplings
 (\ref{LDM}) 
  cancel  in the tree-level diagrams contributing to
 the mass differences.
 
 The independent parameters entering into
 $\Delta M_{D}$ are:
 \be
\sin\beta,~ \sin \gamma^u, ~
(M_H^u)^2= \frac{(M_{H1}M_{H2})^2}{m^2_{\phi^d_H}},~
(M_-^u)^2=\frac{(M_{-1}M_{-2})^2}{m^2_{\phi^d_-}},
\label{param-u}
\ee
where 
they are given, respectively, 
in (\ref{beta}), (\ref{cosgamma}), (\ref{mandb}), and (\ref{M1M2}).
Similarly,  
 \be
\cos\beta,~ \sin \gamma^d, ~
(M_H^d)^2= \frac{(M_{H1}M_{H2})^2}{m^2_{\phi^u_H}},~
(M_-^d)^2=\frac{(M_{-1}M_{-2})^2}{m^2_{\phi^u_-}}
\label{param-d}
\ee
enter into
$\Delta M_{K},
\Delta M_{B_s}$ and $\Delta M_{B_d}$.
 With these remarks in mind, we proceed.
 
\vspace{0.3cm}
\noindent
{\bf D1:} Constraint from $\Delta M_{D}$

\begin{figure}[htb]
\begin{center}\begin{picture}(300,110)(0,45)
\DashLine(130,100)(230,100){3}
\Text(180,80)[]{$\phi_H^u$}
\Text(180,120)[]{$\phi_-^u$}
\ArrowLine(80,150)(130,100)
\Text(100,145)[]{$c_L$}
\Text(100,55)[]{$\overline{u}_R$}
\ArrowLine(130,100)(80,50)
\Text(260,145)[]{$c_R$}
\Text(260,55)[]{$\overline{u}_L$}
\Text(90,100)[]{$Y_{cu}^{uH,-}$}
\Text(270,100)[]{$(Y_{uc}^{uH,-})^*$}
\ArrowLine(280,150)(230,100)
\ArrowLine(230,100)(280,50)
\end{picture}\end{center}
\caption[]{\label{fcnc} The tree-diagram contributing
to $(M_D^{\rm EXTRA})_{12}$. Tree-diagrams
contributing to $M_{K}$ and $M_{B_{d,s}}$ are similar to this diagram.
Leading QCD corrections \cite{Buchalla:1995vs}
will be included, except for $\Delta M_K$.}
\end{figure}
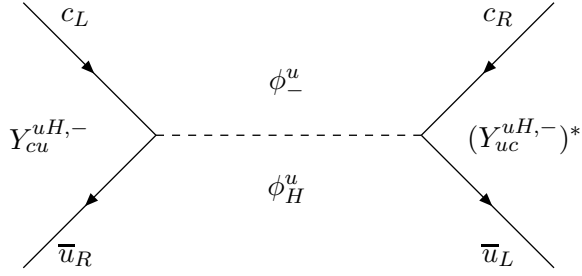

\vspace{0.3cm}
\noindent
The following case is a special case of \cite{Golowich:2007ka},
and we will basically follow their analysis.
An important difference  here is that  the size of 
all the Yukawa couplings is explicitly known.
The tree-level diagrams are shown in Fig.~\ref{fcnc}.
As we can see from Fig.~\ref{fcnc}, only the $\overline{u}_R c_L 
\overline{u}_L c_R $ type operator contributes
to $\Delta M_{D}$ at the tree-level.
The mass difference $\Delta M_{D}$ can then be obtained
from
\be
\Delta M_{D}
&=& 2 \left| (M_D^{\rm SM})_{12}+
(M_D^{\rm EXTRA})_{12}\right |,
 \ee
 where
$(M_D^{\rm SM})_{12}$ is the SM contribution, and
\be
(M_D^{\rm EXTRA})_{12}
  &=&2C_{D}(\mu)
<\overline{D}^0|
\overline{u}_R^\alpha c_L^\alpha 
\overline{u}_L^\beta c_R^\beta |D^0>(\mu),\\
C_{D}(\mu)
&=&   \eta(\mu) \left[  
\frac{Y_{cu}^{uH}(Y_{uc}^{uH})^*}{(M_H^u)^2}+
\frac{Y_{cu}^{u-}(Y_{uc}^{u-})^*}{(M_-^u)^2}\right]
\ee
with the QCD correction $\eta(\mu)$.
The operator 
$\overline{u}_R c_L 
\overline{u}_L c_R $ can mix with
$\overline{u}_L \gamma^\mu c_L 
\overline{u}_R \gamma_\mu c_R$
even at the leading order in QCD in principle
\cite{Ciuchini:1997bw}.  However, if 
$\overline{u}_L \gamma^\mu c_L 
\overline{u}_R \gamma_\mu c_R$
is absent at $\mu=$ some energy,
it will not be induced,
at least in the leading order in QCD.
Note that the values of the Yukawa matrices
(\ref{YuH})-(\ref{Ydm}) are defined at $\mu=M_Z$,
so that there are corrections if $\mu \neq M_Z$.
We here take into account only QCD corrections
because they are most dominant.
The leading-order QCD correction $\eta$ takes the form
\cite{Ciuchini:1997bw}
\be
\eta(\mu_c=2.8\,\mbox{GeV}) &=&
\left[\frac{\alpha_s (m_b)}{\alpha_s (\mu_c)}\right]^{-24/25}
\left[\frac{\alpha_s (m_t)}{\alpha_s (m_b)}\right]^{-24/23}
\left[\frac{\alpha_s (M)}{\alpha_s (m_t)}\right]^{-8/7}
\left[\frac{\alpha_s (M_Z)}{\alpha_s (M)}\right]^{-8/7}\\
&\simeq &
2.3~,
\ee
where we have used the two-loop running of $\alpha_s(\mu)$
with $\alpha_s(M_Z)=0.119$, and the last factor is 
the QCD correction to the Yukawa matrices.
So, the $M$ (which is supposed to be of
the order of the heavy Higgs masses) dependence cancels nicely.
The matrix element in the vacuum saturation approximation is given by \cite{Gabbiani:1996hi}
\be
<\overline{D}^0|
\overline{u}_R^\alpha c_L^\alpha 
\overline{u}_L^\beta c_R^\beta |D^0>(\mu_c=2.8\,\mbox{GeV})
&=&
\frac{1}{4}f_{D}^2 B_D' M_D  
\left(\frac{M_D}{m_c} \right)^2\nn\\
&\simeq & 3.1\times 10^{-2}
~\mbox{GeV}^3,
\label{ucuc}
\ee
where we 
have used the central  values of  the parameters 
\footnote{Since we take here a conservative standpoint
that the extra contribution can be
as large as the experimental value,
we ignore the details of uncertainties.} given
in Table \ref{table1}. 
($m_c(2.8\,\mbox{GeV})=1.0\,\mbox{GeV}$
which corresponds to $m_c(m_c)=1.3\,\mbox{GeV}.$)

\vspace{0.5cm}
\begin{table}
\begin{center}
\begin{tabular}{|c|c||c|c|}
\hline
Input &  &  Input &   \\ \hline
$f_D$ & $(222.6\pm16.7\begin{array}{c}
+2.8\vspace{-3mm}\\-3.4\end{array})\times 10^{-3}$ GeV 
& $B_D' (2.8 \mbox{GeV}) $  & $1.08\pm 0.03$
\\ \hline
$M_D$ & $1.8645\pm 0.0004$ GeV & 
$\tau_D$ & $(410.1\pm 1.5 )\times 10^{-3}$ ps
\\ \hline
$x_D$ & $(5.3-11.7) \times 10^{-3}$  & $f_K$ & $(159.8\pm 1.4 \pm 0.44)\times 10^{-3}$ GeV
\\ \hline
$f_{B_s}$ & 
$\begin{array}{c} \mbox{I:}~0.240\pm 0.040\\
\mbox{II:}~0.245\pm 0.013
\end{array}$ GeV 
& $B_s'(m_b)$ 
& $1.16\pm 0.02\begin{array}{c}
+0.05\vspace{-3mm}\\-0.07\end{array}$
\\ \hline
$f_{B_s}\sqrt{B_s}$ & $\begin{array}{c} \mbox{I:}~0.221\pm 0.046\\
\mbox{II:}~0.227 \pm 0.017
\end{array}$ GeV & 
$\xi $ & $1.24\pm 0.04$
\\ \hline
$f_{B_d}$ & $0.198\pm 0.017$ GeV & $B_d'(m_b)$ & 
$1.15\pm 0.03\begin{array}{c}
+0.05\vspace{-3mm}\\-0.07\end{array}$
\\ \hline
$M_K$ & $0.497648
\pm 0.000022$ GeV & $\Delta M_K^{\rm exp}$ & $(0.5292\pm
0.0009)\times 10^{-2}~\mbox{ps}^{-1}$ 
\\ \hline
$M_{B_s}$ & $5.3661\pm 0.0006$ GeV
 & $\Delta M_{B_s}^{\rm exp}$ & $17.77\pm 0.10\pm 0.07
 ~\mbox{ps}^{-1}$
\\ \hline
$M_{B_d}$ & $5.27950\pm 0.00033$ GeV 
& $\Delta M_{B_d}^{\rm exp}$ & 
$0.507\pm 0.005 ~\mbox{ps}^{-1}$ 
\\ \hline
$m_u (2 \mbox{GeV})$ & $(3\pm1)\times 10^{-3}$
 GeV & $m_c(m_c)$ & $1.30\pm 0.05$ GeV
\\ \hline
$m_d(2 \mbox{GeV})$ & $(6.0\pm 1.5)\times 10^{-3}$ GeV 
& $m_s (2 \mbox{GeV})$ & $0.10\pm0.02$ GeV
\\ \hline
$m_d(m_b)$ & $(5.1\pm 1.3)
\times 10^{-3}$ GeV & $m_s (m_b)$ & $0.085\pm 0.017$ GeV
\\ \hline
$m_t(m_t) $ & $163.8\pm 2.0$ GeV 
& $m_b(m_b)$ & $4.22\pm 0.08$ GeV
\\ \hline
\end{tabular}
\caption{ \footnotesize{Parameter values used in the text
(see also Ref.~\cite{Bona:2006ah}).
$f_D$ is taken from \cite{Artuso:2005ym}, and
we use $B_D'$  and $x_D$ of \cite{Ciuchini:2007cw}
and \cite{HFAG}, respectively.
$M_D, \tau_D, f_K, M_K, \Delta M_K^{\rm exp}, M_{B_s}, M_{B_d},
\Delta M_{B_d}^{\rm exp}$ are from 
\cite{Yao:2006px}.
$f_{B_s}$ (I) and $f_{B_s}\sqrt{B_s}$ (I)
are the conservative sets of \cite{Lenz:2006hd}, and 
$f_{B_s}\sqrt{B_s}$ (II) is found in
\cite{Dalgic:2006gp}, while
$f_{B_s}$ (II) and $\xi$ are taken from   \cite{Tantalo:2007ai}, and
$f_{B_d}$ is obtained from $f_{B_s}/\xi$.
(See \cite{DellaMorte:2007ny} 
for a more conservative estimate of $\xi$,
and references therein.)
$B_s'$ and $B_d'$ are found in
\cite{Becirevic:2001xt}.
$\Delta M_{B_s}^{\rm exp}$ is from
\cite{Abulencia:2006psa}.
$m_u (2 \mbox{GeV})$ and $m_d (2 \mbox{GeV})$ are from 
\cite{Yao:2006px}, while the mass values of the other quarks
are taken from \cite{Lenz:2006hd},
in which the relevant references are given. }}
\label{table1}
\end{center}
\end{table}

\begin{figure}[htb]
\includegraphics*[width=0.8\textwidth]{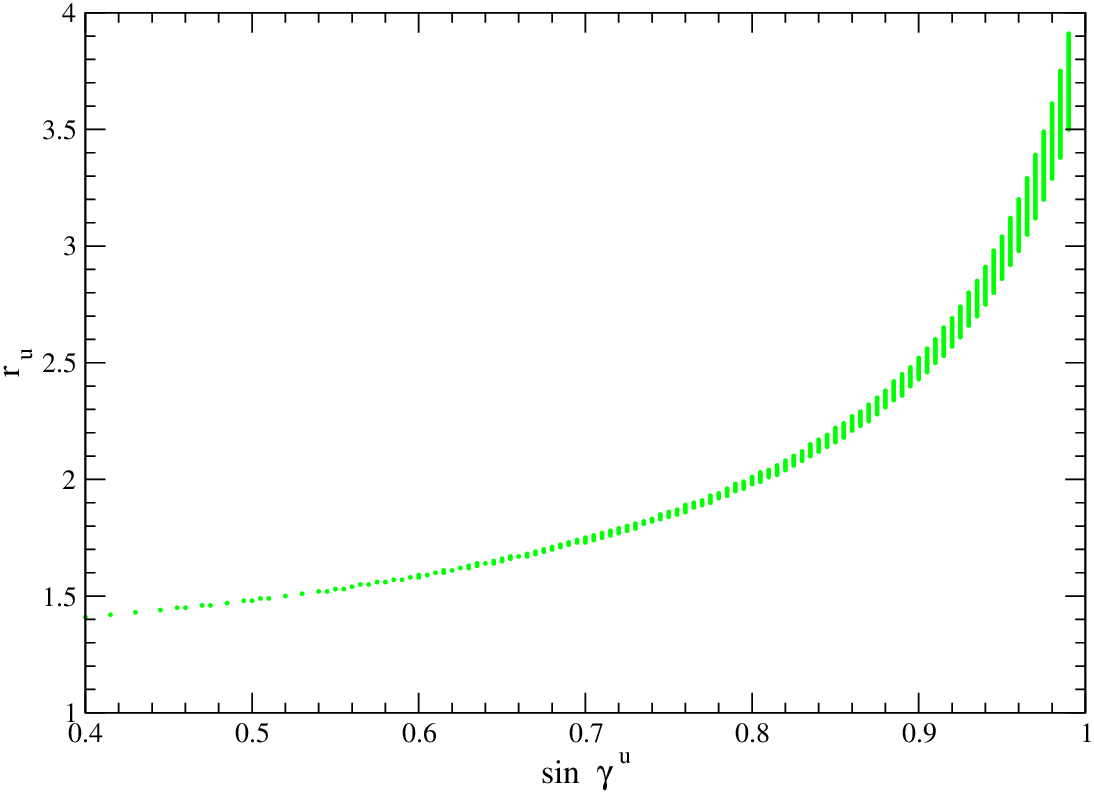}
\caption{\label{sgu-ru}\footnotesize
The  region in the $\sin\gamma^u-r_u$ plane,
in which the constraint (\ref{constD}) coming from
$\Delta M_D$ is satisfied for $\sin\beta M_H^u=
2$ TeV, where $r_u, \sin\gamma^u$ and 
$M_H^u$ are defined in (\ref{r-u}),  (\ref{cosgamma}) and
 (\ref{param-u}), respectively.}
\end{figure}

Clearly, the larger 
$(M_H^u)^2$ and 
$(M_-^u)^2$ are, the smaller are the extra contributions.
Here we are interested in the minimal values of 
$(M_H^d)^2$ and 
$(M_-^d)^2$, which are consistent with the observations.
We find that the Wilson coefficient $C_D$ becomes
\be
C_D(\mu_c) &=& \frac{\eta(\mu_c)}{\sin^2\beta }
\left[  \frac{1\, \mbox{TeV}}{M_H^u}
\right]^2~\times 10^{-11}
\nn\\
&\times & \left(\frac{1.772 }{r_u^2 \sin^2\gamma^u}
-\frac{1.037 }{ \tan^2\gamma^u}-0.115
+5.5\times 10^{-5} \tan^2\gamma^u
\right)~\mbox{GeV}^{-2},
\label{CD}
\ee
where
\be
r_u &=&\frac{M_-^u}{M_H^u}=
\left(\frac{M_{-1} M_{-2}}{M_{H1} M_{H2}}\right)
~\left(\frac{m_{\phi_H^d}}{m_{\phi_-^d}}\right),
\label{r-u}
\ee
and the mass parameters are defined in (\ref{param-u}).
If each term in (\ref{CD}) should satisfy the constraint, 
\be
|\Delta M_D^{\rm EXTRA}| &=&
2\left| (M_D^{\rm EXTRA})_{12}\right| <
 \Delta M_D^{\rm exp}=x_D /\tau_D
  \simeq1.4 \times 10^{-14}~\mbox{GeV},
 \ee
one finds that 
$\sin\beta M_H^u \gsim 17$ TeV and   
 $\sin\beta M_-^u \gsim 22$ TeV should be satisfied.
We, however, observe  that 
the terms in (\ref{CD}) can cancel each other, so that no lower bounds on
$M_H^u$  and    $M_-^u$ can be obtained.
In Fig.~\ref{sgu-ru} we show the  region in
the $\sin\gamma^u-r_u$ plane
for $\sin\beta  M_H^u= 2$ TeV,
in which 
$|\Delta M_D^{\rm EXTRA}| $ is
smaller than the smallest
$ \Delta M_D^{\rm exp}$, i.e.,
\be
|\Delta M_D^{\rm EXTRA}|  &<&
8\times 10^{-15}  ~\mbox{GeV}.
\label{constD}
\ee
We see from Fig.~\ref{sgu-ru} that to satisfy the constraint
(\ref{constD}),
we have to fine tune $r_u$ and $\sin\gamma^u$ even
 for $\sin\beta  M_H^u= 2$ TeV.

The neutral Higgs bosons in question can induce
 processes such as
$D^0 \to e^+~e^-$ and $D^0 \to \mu^+~\mu^-$
which are strongly suppressed.
The experimental upper bounds of the branching ratios are  smaller than 
$O(10^{-6})$. From a rough estimate we find
that $M_-^u,M_H^u > M_Z$ is more than sufficient to 
suppress these processes.
So, in principle,  $M_-^u,M_H^u $ could be light,
although one needs an extreme fine tuning
between $r_u$ and $\sin\gamma^u$.

\vspace{0.3cm}
\noindent
{\bf D2:} Constraint from $\Delta M_{K}$

\vspace{0.3cm}
\noindent
As in the case of $\Delta M_{D}$, the interaction  Lagrangian generates 
only one type of the $\Delta S=2$ operator at the tree level.
So, the relevant matrix element is
\be
<\overline{K}^0|
\overline{s}_R^\alpha d_L^\alpha 
\overline{s}_L^\beta d_R^\beta |K^0>
&=&
\frac{1}{4}f_{B_K}^2 B_K' M_K  
\left(\frac{M_K}{m_s+m_d} \right)^2
\label{slsl}\\
&\simeq & 0.28~\mbox{GeV}^3,\nn
\ee
where we have used the central values of
the parameters given in Table \ref{table1}.
(As in the case of $\Delta M_D$ we 
we ignore the details of uncertainties involved in 
$\Delta M_D$.)
As far as we understand, there is no reliable calculation
of $B_K'$ for the present case (\ref{slsl}), \footnote{See 
\cite{Aoki:2005ga} for a lattice calculation of 
$B_K'$ of the present case, and also comments of
\cite{Becirevic:2004qd}.}  and so we
assume that $B_K'=1$.
Correspondingly, we do not take into account QCD corrections
for the present case.

The tree-level coefficient  is given by
\be
C_K &=&
\left[  
\frac{Y_{ds}^{dH}(Y_{sd}^{dH})^*}{(M_H^d)^2}+
\frac{Y_{ds}^{d-}(Y_{sd}^{d-})^*}{(M_-^d)^2}\right]\nn\\
&=&
\frac{1}{\cos^2\beta }
\left[  \frac{1\, \mbox{TeV}}{M_H^d}
\right]^2~\times 10^{-14}
\nn\\
& \times & \left(\frac{5.617 }{r_d^2 \sin^2\gamma^d}
-\frac{0.514 }{ \tan^2\gamma^d}-0.539
+0.224\tan^2\gamma^d
\right)\, \mbox{GeV}^{-2},
\ee
where
\be
r_d &=&\frac{M_-^d}{M_H^d}=
\left(\frac{M_{-1} M_{-2}}{M_{H1} M_{H2}}\right)
~\left(\frac{m_{\phi_H^d}}{m_{\phi_-^d}}\right).
\label{r-d}
\ee
\begin{figure}[htb]
\includegraphics*[width=0.8\textwidth]{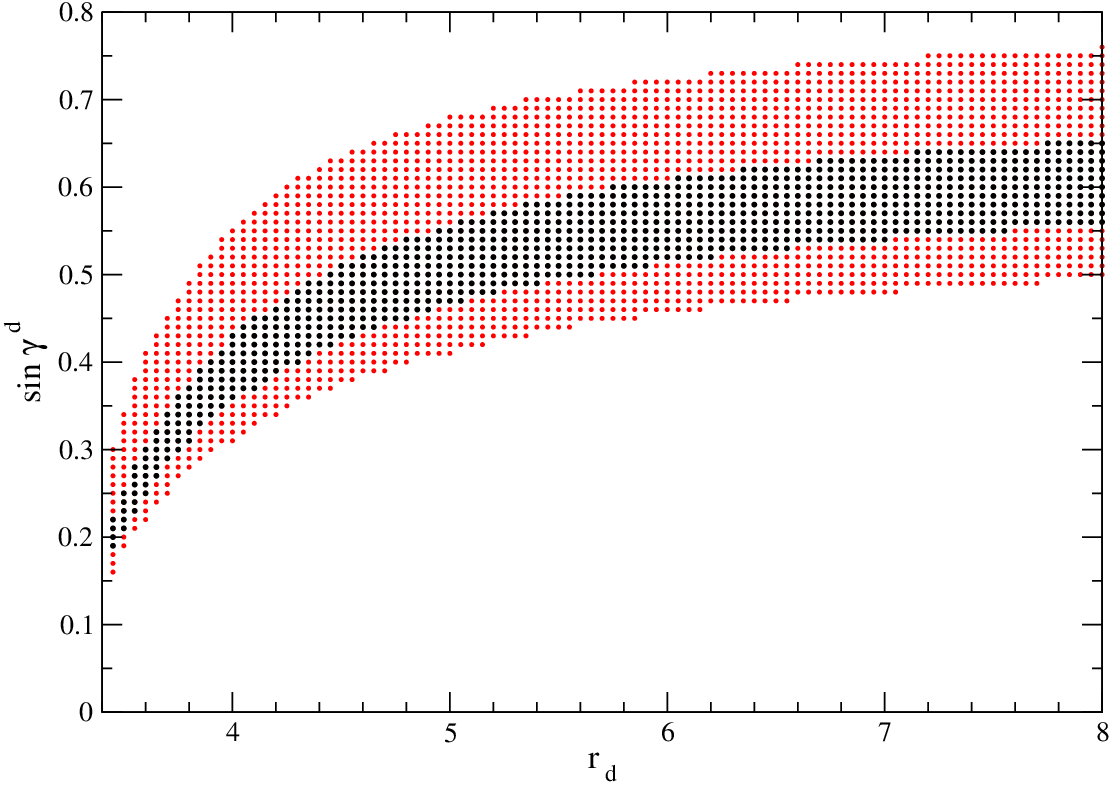}
\caption{\label{rd-sgd-K}\footnotesize
The  region in
the $r_d-\sin\gamma^d$ plane
for $\cos\beta  M_H^d= 0.5$ TeV (red (dark grey))
and  $0.3$ TeV (black),
in which 
$|\Delta M_K^{\rm EXTRA}| < 
\Delta M_{K}^{\rm exp}$ is satisfied. 
$r_d$  and $\sin \gamma^d$
are defined in (\ref{r-d}) and (\ref{cosgamma}), respectively.
}
\end{figure}
In Fig.~\ref{rd-sgd-K} we show the region in the 
$r_d-\sin\gamma^d$ plane   in which 
\be
\Delta M_{K}& =&
 2\times 0.28\times C_K\,\mbox{GeV} < \Delta M_{K}^{\rm exp}
\simeq 3.49 \times 10^{-15}~\mbox{GeV}
\label{constK}
\ee
is satisfied.

\vspace{0.3cm}
\noindent
{\bf D3:}  Constraint from $\Delta M_{B_s},\Delta M_{B_d}$

\vspace{0.3cm}
\noindent
As in the previous cases, the mass differences can be obtained from
\be
\Delta M_{B_{s,d}} &=& 2\left| <\overline{B}^0|
~(M_{s,d}^{\rm SM})_{12}
+(M_{s,d}^{\rm EXTRA})_{12}~ | B^0>\right|.
\ee
 The SM contributions to  
$\Delta M_{B_s},\Delta M_{B_d}$ are well controlled 
up to the numerical uncertainty 
in the decay constants. Here following \cite{Lenz:2006hd},
which is based on the NLO-QCD calculations in
  Refs.~\cite{Beneke:2003az}
  and   \cite{Beneke:1998sy},
we consider two sets of  the uncertainties for the $B$ system,
I and II, 
as one can see in Table \ref{table1}.
Since the uncertainties in the decay constants are
much larger than those of other quantities,
we assume
that
\be
f_{B_s}\sqrt{B_s} &=&\left\{ \begin{array}{c} 
0.221\pm 0.046~\mbox{for the parameter set I}\\
0.227 \pm 0.017~\mbox{for the parameter set II}\end{array}\right.~,\\
f_{B_d}\sqrt{B_d} &=&\left\{ \begin{array}{c} 
0.181\pm 0.044~\mbox{for the parameter set I}\\
0.184 \pm 0.017~\mbox{for the parameter set II}\end{array}\right.
\label{uncertainty}
\ee
are the only uncertainties for the SM model
contributions $M_{s,d}^{\rm SM}$ , where
$f_{B_d}\sqrt{B_d}$ is obtained from 
$\xi=f_{B_s}\sqrt{B_s}/f_{B_d}\sqrt{B_d}$.
To simplify the situation further, we
 assume that this is also true for the extra contributions
$M_{s,d}^{\rm EXTRA}$.

To calculate $(M_{s,d}^{\rm SM})_{12}$
we  use the parameter values (\ref{predQ}) which are predicted in
the present model:
\be
|V_{\rm CKM}|_{us} &=&0.2266,~
|V_{\rm CKM}|_{ub} =0.00362,~|V_{\rm CKM}|_{cb} =0.0417,~
\phi_3(\gamma)=1.107.
\ee
Then we follow the calculation of \cite{Lenz:2006hd} and obtain:
\be
2(M_{B_s}^{\rm SM})_{12}&=&
2\left| (\bar{M}_s^{\rm SM})_{12}\right|(1\pm \delta_s)
\exp i \phi_s\nn\\
&=&\left\{~\begin{array}{c}
19.5(1\pm 0.46)\exp (-i 0.0035)\\
20.6(1\pm 0.16)\exp (-i 0.0035)
\end{array}\right.
~\mbox{ps}^{-1}~\mbox{for}~
\left\{~\begin{array}{c}\mbox{I}\\
\mbox{II}\end{array}\right.~,
\label{phis}\\
2(M_{B_s}^{\rm SM})_{12}
&=&2\left|(\bar{M}_{d}^{\rm SM})_{12}\right|(1\pm \delta_d)
\exp i \phi_d\nn\\
&=&\left\{~\begin{array}{c}
0.56(1\pm 0.55)\exp (i 0.77)\\
0.59(1\pm 0.19)\exp (i 0.77)
\end{array}\right.
~\mbox{ps}^{-1}~\mbox{for}~
\left\{~\begin{array}{c}\mbox{I}\\
\mbox{II}\end{array}\right.~,
\label{phid}
\ee
where $(\bar{M}_{s,d}^{\rm SM})_{12}$ are the SM
contributions which are obtained with the central values of 
$f_{B_s}\sqrt{B_s},\xi, M_{B_{s,d}}$ and
the quark masses given
\footnote{The model does not predict
the absolute scale for the quark masses.
If we use the mass ratio given in (\ref{predQ}),
we obtain a slightly smaller value for
$m_b(m_b)$ (while we obtain the same value
for $m_c(m_c)$). This difference  has only a negligible effect on 
the SM contributions.
}
in  \ref{table1}  and $\alpha_s(M_Z)=0.119$, and
$\delta_s$ and $\delta_d$
 correspond to the uncertainties 
in $f_{B_s}\sqrt{B_s}$ and $f_{B_d}\sqrt{B_d}$
given in (\ref{uncertainty}), respectively.
As we can see from Table \ref{table1}, the SM values are
slightly larger than the experimental values.

\begin{figure}[htb]
\includegraphics*[width=0.8\textwidth]{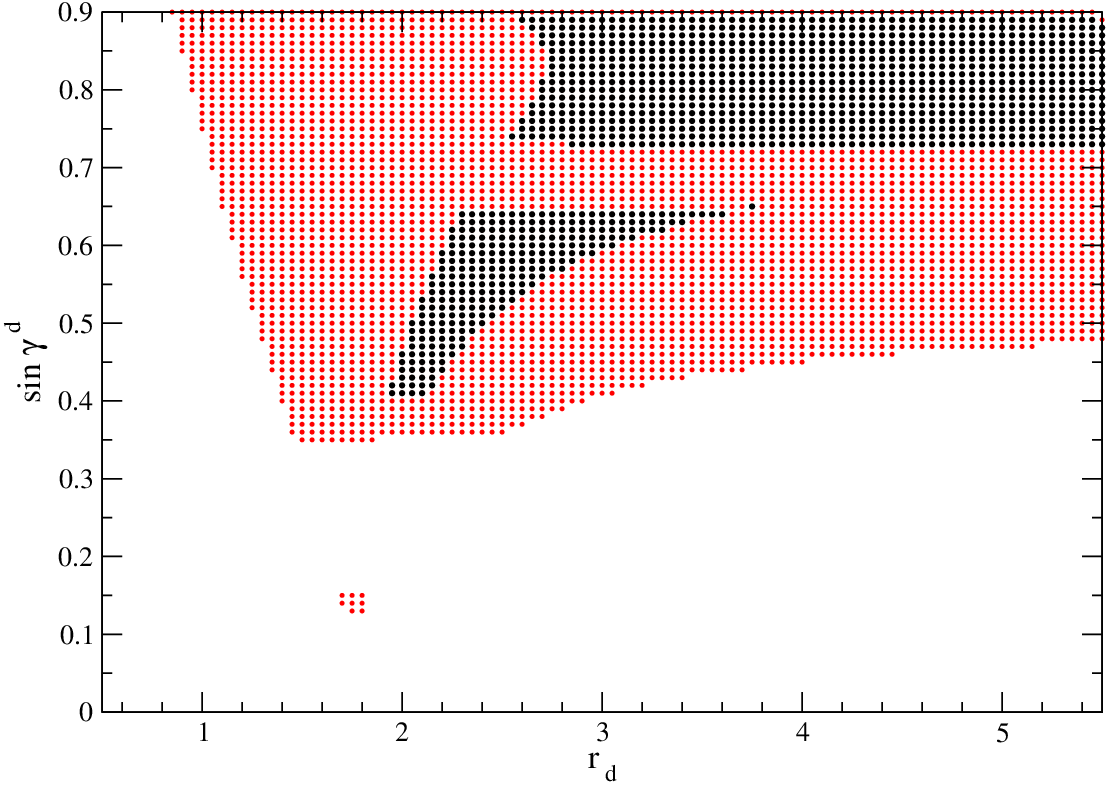}
\caption{\label{rd-sgd-ds}\footnotesize
The  allowed region 
for the parameter set  I with $\cos\beta  M_H^d= 0.50$ 
(black) and 
$1.5$ (red (dark grey)) TeV
in which the constraints (\ref{constBS}) and 
(\ref{constBD}) are simultaneously  satisfied. 
$r_d$  and $\sin \gamma^d$
are defined in (\ref{r-d}) and (\ref{cosgamma}), respectively.
Two sets of values I and II are given in Table \ref{table1}.
}
\end{figure}
\begin{figure}[htb]
\includegraphics*[width=0.8\textwidth]{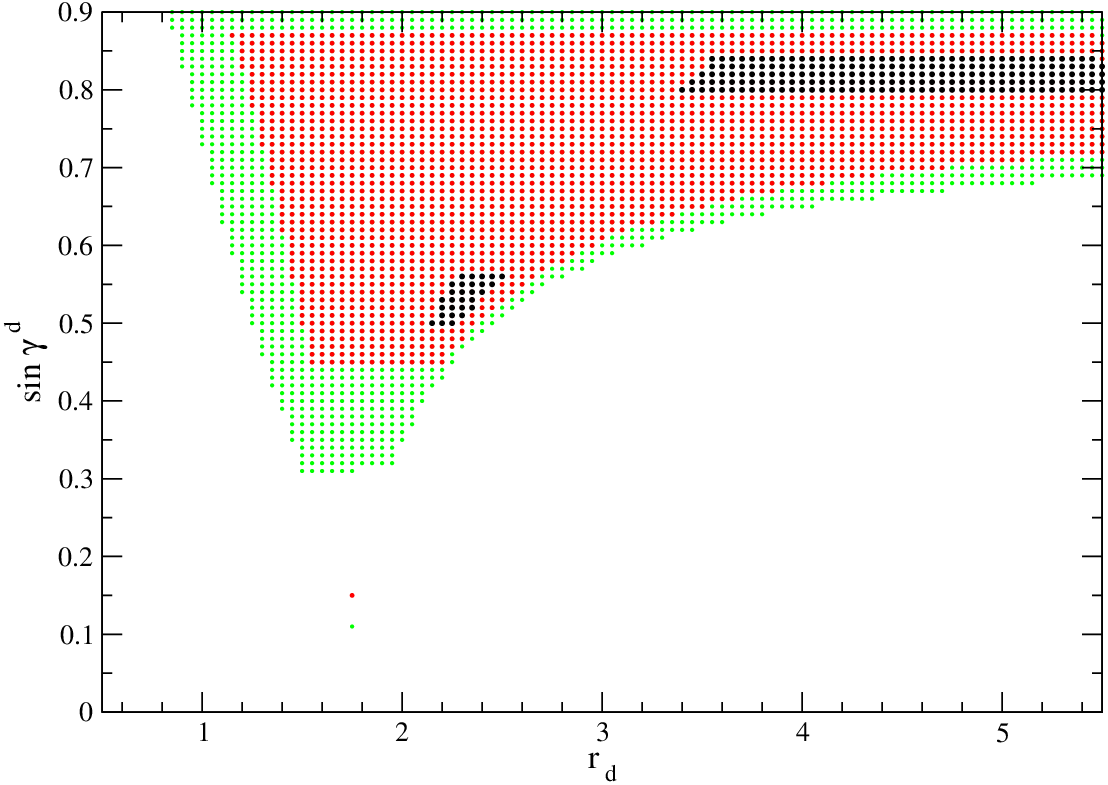}
\caption{\label{rd-sgd-ds2}\footnotesize
The  same as Fig.~\ref{rd-sgd-ds} for the parameter set II with
$\cos\beta  M_H^d= 0.50$ (black),
$1.5$ (red (dark grey)) and $2$ (green (grey)) TeV .}
\end{figure}
\begin{figure}[htb]
\includegraphics*[width=0.8\textwidth]{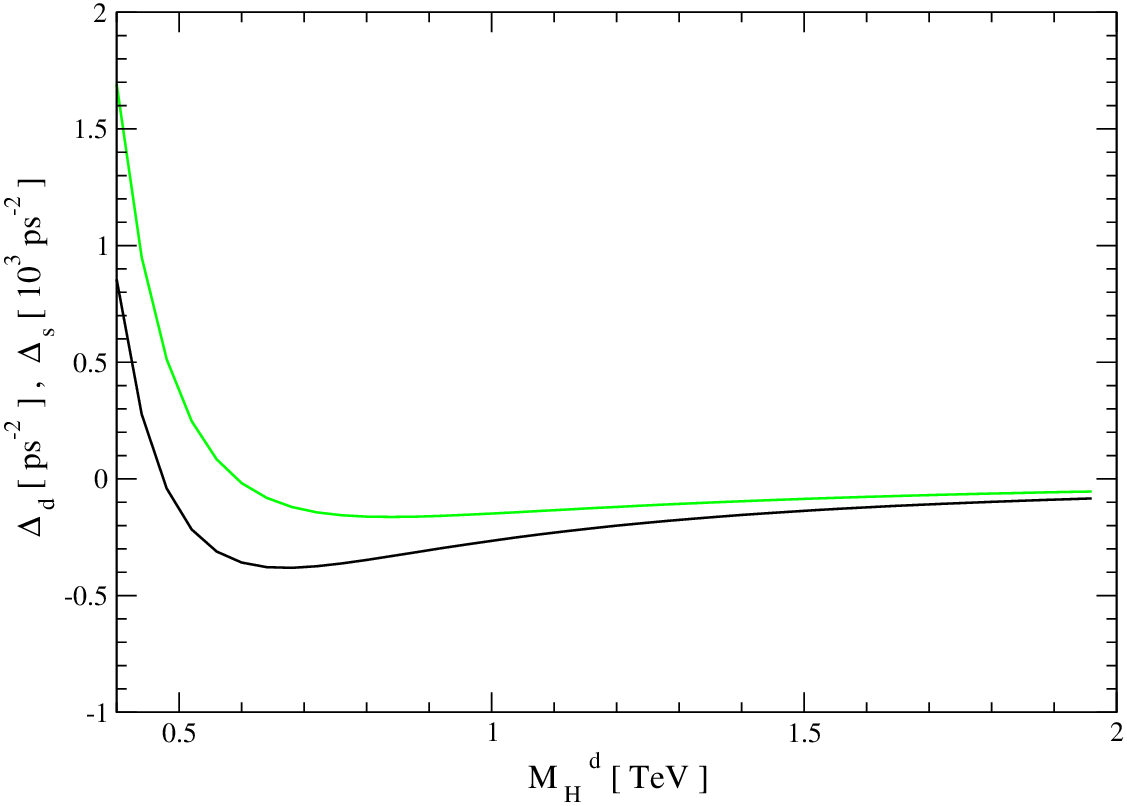}
\caption{\label{delta-sd1}\footnotesize
$\Delta_s$ (green (grey)) and $\Delta_d$  (black) for 
the parameter set I with
$r_d=  3$ and $\sin\gamma^d=  0.8$, where they are defined in
(\ref{deltasd}). This graph explains why the allowed 
region in the $r_d-\sin\gamma^d$ plane
first shrinks and then extends as $M_H^d$ increases.
For the the parameter set  II we obtain a similar result.
}
\end{figure}
\begin{figure}[htb]
\includegraphics*[width=0.8\textwidth]{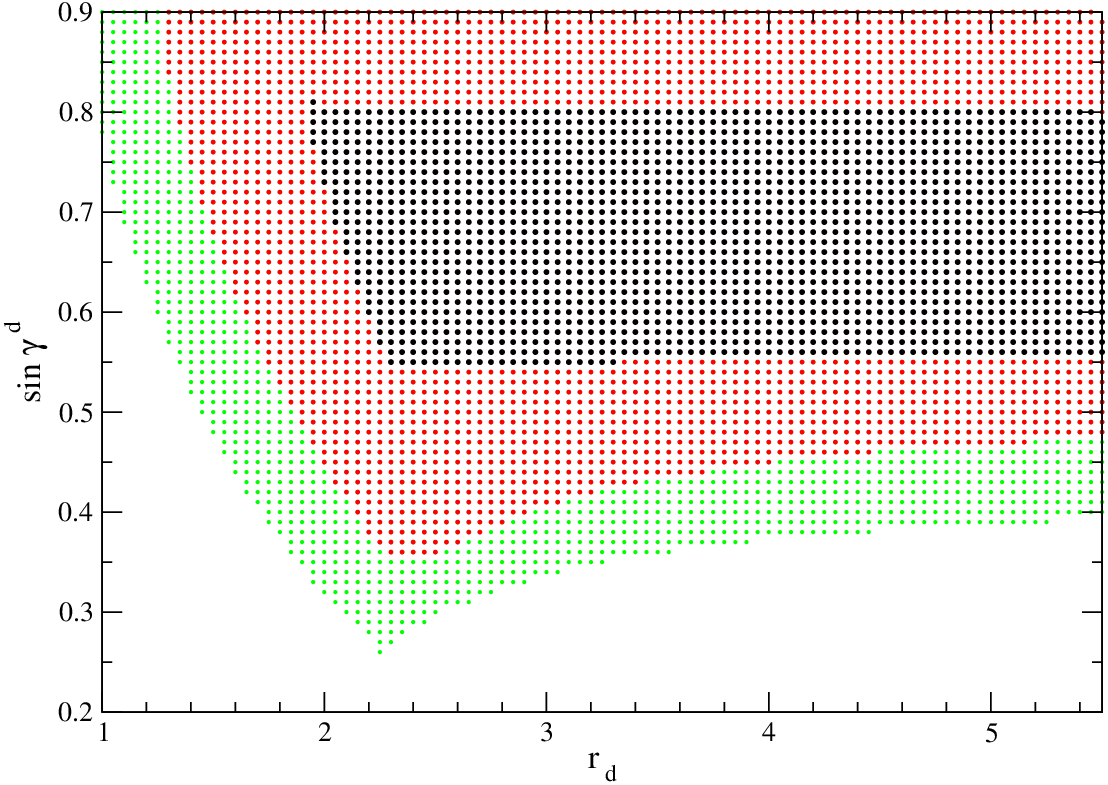}
\caption{\label{rd-sgd-dsk}\footnotesize
The region in which the constraints (\ref{constK}),
(\ref{constBS}) and (\ref{constBD}) 
coming from $\Delta M_K, \Delta M_{B_{s,d}}$  are satisfied
for the parameter set I with $M_H^d =1.1$ (black), $M_H^d =1.5$ (red (dark grey)) and $2$  (green (grey)) TeV.
$r_d$  and $\sin \gamma^d$
are defined in (\ref{r-d}) and (\ref{cosgamma}), respectively.}
\end{figure}
\begin{figure}[htb]
\includegraphics*[width=0.8\textwidth]{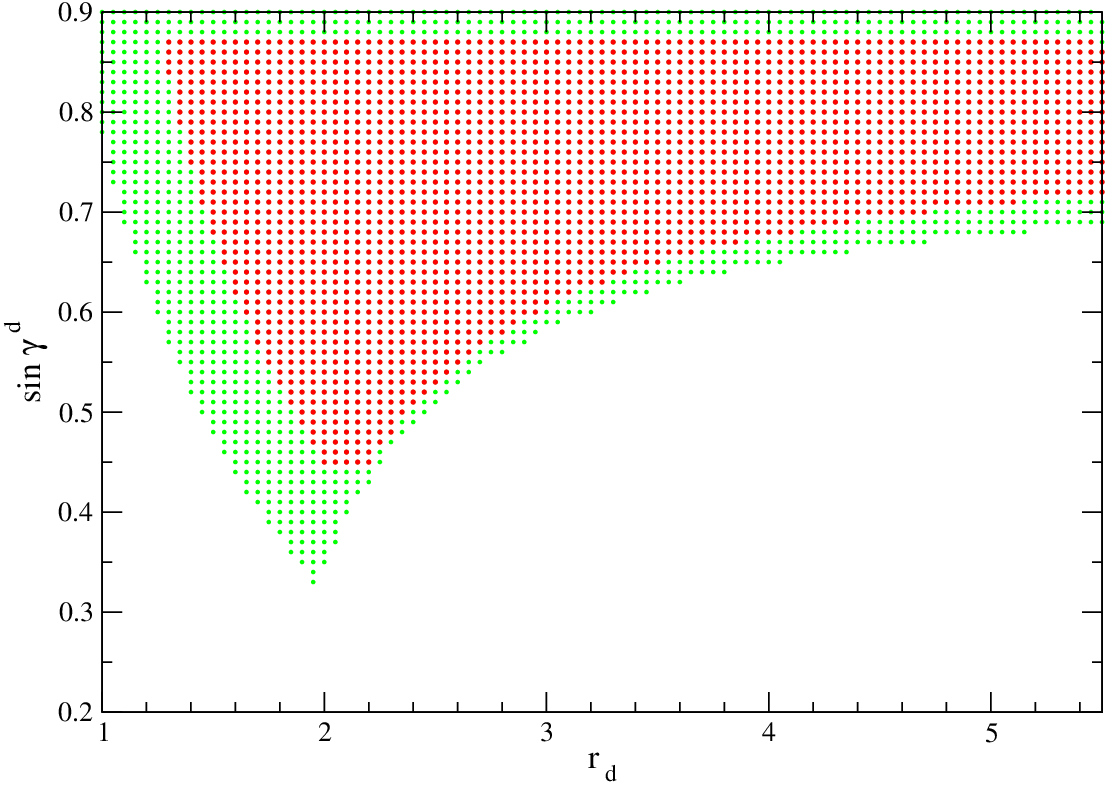}
\caption{\label{rd-sgd-dsk2}\footnotesize
The same as Fig.~\ref{rd-sgd-dsk} for the parameter set II with
$M_H^d =1.5$ (red (dark grey)) and $2$ (green (grey))
TeV.}
\end{figure}
\begin{figure}[htb]
\includegraphics*[width=0.8\textwidth]{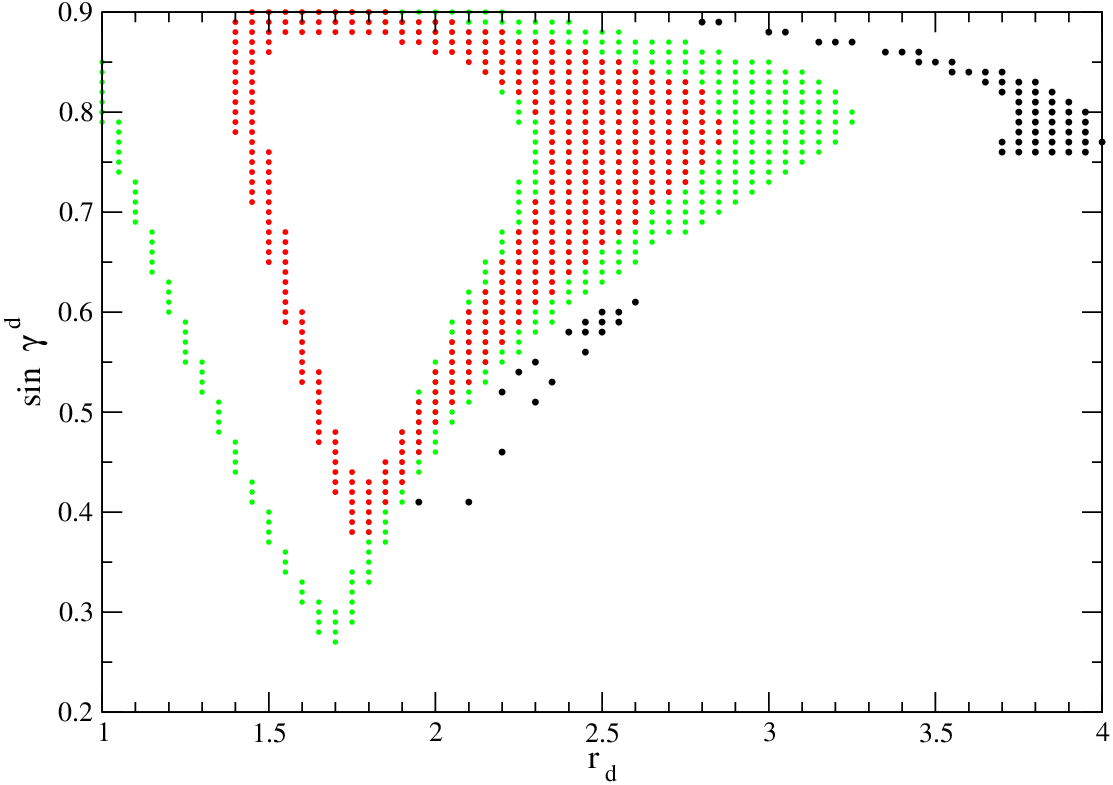}
\caption{\label{rd-sgd-dsr}\footnotesize
The  allowed region for the parameter set I with
 $\cos\beta  M_H^d= 0.50$ (black),
$1.5$ (red (dark grey)) and $ 2$ (green (grey)) TeV,
in which the constraints (\ref{constBS}),
(\ref{constBD}) and
(\ref{constR}) coming from 
$\Delta M_{B_{s,d}}$ and $ \Delta M_{B_s}/\Delta M_{B_d}$
are simultaneously  satisfied. 
}
\end{figure}
\begin{figure}[htb]
\includegraphics*[width=0.8\textwidth]{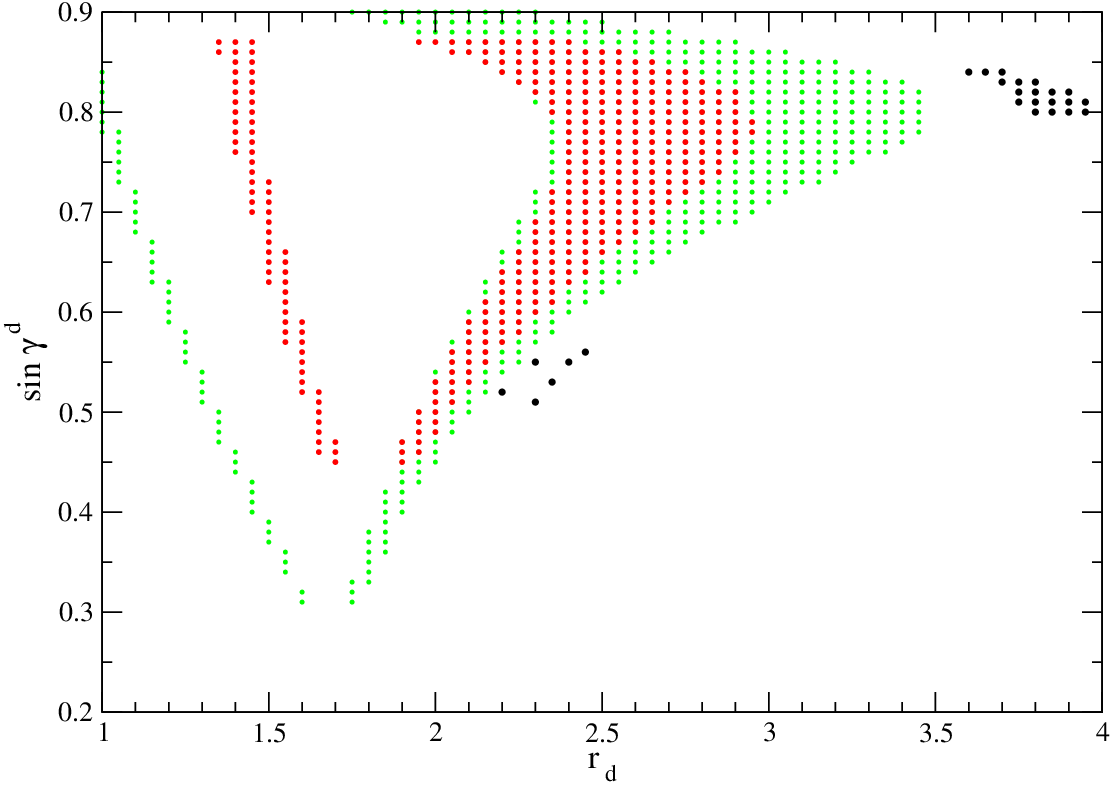}
\caption{\label{rd-sgd-dsr2}\footnotesize
The same as Fig.~\ref{rd-sgd-dsr} for the parameter set II.
Two set of values I and II are given in Table \ref{table1}.}
\end{figure}

As for the extra contributions, only the  matrix elements 
\be
<\overline{B_s}^0|
\overline{b}_R^\alpha s_L^\alpha 
\overline{b}_L^\beta s_R^\beta |B_s^0>
&=&
\frac{1}{4}f_{B_s}^2 B_s' M_{B_s}
\left(\frac{M_{B_s}}{m_b+m_s} \right)^2\nn\\
&\simeq &\left\{ \begin{array}{c}0.29~\mbox{(I)}\\
0.30~\mbox{(II)} \end{array}~\mbox{GeV}^3\right.
\ee
and
\be
<\overline{B_d}^0|
\overline{b}_R^\alpha d_L^\alpha 
\overline{b}_L^\beta d_R^\beta |B_d^0>
&=&
\frac{1}{4}f_{B_d}^2 B_d' M_{B_d} 
\left(\frac{M_{B_d}}{m_b+m_d} \right)^2\nn\\
&\simeq &
0.18~\mbox{(I,II)}~\mbox{GeV}^3
\ee
are relevant for $\Delta M_{B_s},\Delta M_{B_d}$,
where the tree-level diagrams similar to
Fig.~\ref{fcnc} contribute to these
mass differences,
and we have used the central values of the
parameters in Table \ref{table1}. The leading order Wilson 
coefficients are
\be
C_{B_s} &=&  \eta_B(m_b) 
\frac{1}{\cos^2\beta }
\left[  \frac{1\, \mbox{TeV}}{M_H^d}
\right]^2~\times 10^{-12}
\nn\\
&\times &\left(\frac{0.197 }{r_d^2 \sin^2\gamma^d}
-\frac{2.025 }{ \tan^2\gamma^d}-4.463
-2.459\tan^2\gamma^d
\right)~\mbox{GeV}^{-2},\\
C_{B_d} &=&\eta_B(m_b) \frac{1}{\cos^2\beta }
\left[  \frac{1\, \mbox{TeV}}{M_H^d}
\right]^2~\times 10^{-12}
\nn\\
&\times &\left(-\frac{3.521 }{r_d^2 \sin^2\gamma^d}
+\frac{1.148 }{ \tan^2\gamma^d}-0.780
+0.127\tan^2\gamma^d
\right)~\mbox{GeV}^{-2}~,
\ee
where
\be
\eta_B(m_b=4.22\,\mbox{GeV}) &=&
\left[\frac{\alpha_s (m_t)}{\alpha_s (m_b)}\right]^{-24/23}
\left[\frac{\alpha_s (M)}{\alpha_s (m_t)}\right]^{-8/7}
\left[\frac{\alpha_s (M_Z)}{\alpha_s (M)}\right]^{-8/7}\nn\\
&\simeq &
2.0~.
\ee
Then we require that
\be
\Delta M_{B_s} &=& \Delta M_{B_s}^{\rm exp}=
2\left| (M_{B_s}^{\rm SM})_{12}+0.41\times C_{B_s}\right|=
17.77~\mbox{ps}^{-1}=1.17\times 10^{-11}~\mbox{GeV}~,
\label{constBS}\\
\Delta M_{B_d} &=& \Delta M_{B_d}^{\rm exp}=
2\left| (M_{B_d}^{\rm SM})_{12}+0.25\times C_{B_d}\right|=
0.507~\mbox{ps}^{-1}=3.34\times 10^{-13}~\mbox{GeV}~.
\label{constBD}
\ee
Note that according to our assumption
the uncertainties factorize
as $(1\pm \delta_{s,d})[\,(\bar{M}_{s,d}^{\rm SM})_{12}
+(M_{s,d}^{\rm EXTRA})_{12}\,]$,
where $\bar{M}_{s,d}$ are the central values
and  $\delta_{s(d)}$ are given
in (\ref{phis}) and (\ref{phid}).
In Fig.  \ref{rd-sgd-ds} and  \ref{rd-sgd-ds2} we show the allowed region in 
the $r_d-\sin\gamma^d$ plane for the parameter sets I and II, 
respectively, in which 
(\ref{constBS}) and (\ref{constBD}) are satisfied.
We find that (\ref{constBS}) and (\ref{constBD})
can be simultaneously satisfied even for small $M_H^d \gsim 0.50$ TeV.
The allowed region shrinks as $M_H^d$ increases.
At $M_H^d  =1$ TeV the allowed region is very small.
But a wide allowed region exists for $M_H^d  =2$ TeV.
The reason that the allowed region first decreases
and then increases as $M_H^d$ increases starting from
$\simeq 0.50$ TeV is the following.
The constraint (\ref{constBS}) and
 and (\ref{constBD})can be written
as
\be
[\Delta M_{B_{s,d}}^{\rm exp}]^2(1+\delta_s)^{-2}
&\leq &  \Delta_{s,d} +4 [(\bar{M}_s^{\rm SM})_{12}]^2 \leq  
[\Delta M_{B_{s,d}}^{\rm exp}]^2(1-\delta_s)^{-2}~,
\label{constSD}
\ee
where
\be
 \Delta_{s,d} &=& 4[(M_{s,d}^{\rm EXTRA})_{12}]^2
+8\cos\phi_{s,d} (\bar{M}_{s,d}^{\rm SM})_{12}
(M_{s,d}^{\rm EXTRA})_{12}~,
\label{deltasd}
\ee
and $(\bar{M}_{s,d}^{\rm SM})_{12}$
and $\phi_{s,d}$ are given in (\ref{phis}) and (\ref{phid}).
For a large $M_H^d$ the second term
of $\Delta_{s,d}$ is dominant. However, for a small $M_H^d$,
two terms can become of the same order, and
since $(M_{B_s}^{\rm EXTRA})_{12}$
and $(M_{B_d}^{\rm EXTRA})_{12}$
can simultaneously become negative, these two terms can cancel
each other, so that the both constraints (\ref{constSD})
for $\Delta M_{B_s}$ and $\Delta M_{B_d}$
can be simultaneously satisfied.
In Fig.~\ref{delta-sd1} we show 
$\Delta_s$ (red) and $\Delta_d$ (blue) for the parameter set I
as a function of $M_H^d$ for
$r_d=3$ and $\sin\gamma^d=0.8$, where we vary
$M_H^d$ from $0.4$ TeV to $2$ TeV.
We see from the figure that
$\Delta_s$  and $\Delta_d$ decrease
as $M_H^d$ increases for $M_H^d \lsim 0.6$ TeV.
Note that the constraint from $\Delta M_{B_d}$
(\ref{constBD}) is stronger than 
that from  $\Delta M_{B_s}$
(\ref{constBS}).
In this region the constraint from 
$\Delta_K$ is not satisfied. But if we relax
the constraint (because non-perturbative
contributions to $\Delta_K$ suffer from large uncertainties)
to $\Delta M_K^{\rm EXTRA}
< 2 \Delta M_K^{\rm exp}$,
then it is satisfied.

Next we consider the region in which all the three
constraints  (\ref{constK}),
(\ref{constBS}) and (\ref{constBD}) are satisfied.
We find that the small $M_H^d$ region 
in Fig.~\ref{rd-sgd-ds} and \ref{rd-sgd-ds2}
disappears, and that $M_H^d \gsim 1.0$ (I) and $ 1.3$ (II) TeV
have to be satisfied.
In Fig.~ \ref{rd-sgd-dsk} and
\ref{rd-sgd-dsk2} we show the allowed region in which
all the constraints  (\ref{constK}),
(\ref{constBS}) and (\ref{constBD}) are satisfied
for $M_H^d = 1.5$ TeV (blue)
$M_H^d = 2$ TeV (green).

\vspace{0.3cm}
\noindent
{\bf D4:} Constraint from $\Delta M_{B_s}/\Delta M_{B_d}$

This ratio is important to determine experimentally
$|V_{td}/V_{ts}|$. This is true only if there is no 
 other contribution than the SM ones.
 In the presence of the extra neutral Higgs bosons,
 the situation changes.
 Here we ask ourselves how heavy the extra 
extra neutral Higgs bosons should be, or
where the allowed region in the 
$r_d-\sin\gamma^d$ plane for a given $\cos\beta M_H^d$ is, such that
the determination of 
$|V_{td}/V_{ts}|$  from the ratio 
$\Delta M_{B_s}/\Delta M_{B_d}$
is not influenced.

The largest theoretical uncertainty in the mass ratio is contained in
$\xi=f_{B_s}\sqrt{B_s}/f_{B_d}\sqrt{B_d}
=1.24\pm 0.04$ (see Table \ref{table1}),
that is, $3.2\%$ uncertainty,
which is lager than the experimental ones.
Accordingly, 
we require that
the theoretical value of $\Delta M_{B_s}/\Delta M_{B_d}$
 should be equal to the experimental central value
 $35.05$ within an error of $5\%$
 (the mass ratio is proportional to $\xi^2$),
 i.e. \be
 \Delta M_{B_s}/\Delta M_{B_d} &=& 35.05(1 \pm 0.05)~.
 \label{constR}
 \ee
We require that (\ref{constBS}), (\ref{constBD}) 
and (\ref{constR}) are simultaneously satisfied.
The allowed region is shown in
 Fig.~\ref{rd-sgd-dsr} and \ref{rd-sgd-dsr2}
 for $M_H^d=0.50 ~\mbox{(red)}, 1.5 ~
 \mbox{(blue)} $ and $2~ \mbox{(green)}$ TeV.
 We see that the small $M_H^d$ region of 
  Fig.~\ref{rd-sgd-ds} and  \ref{rd-sgd-ds2} is still there.
 We also find that
$M_H^d \gsim1.1~(1.3)$ for the parameter set I (II) TeV
or $0.39~\mbox{TeV}~\lsim M_H^d \lsim 0.65$ 
( $0.45~\mbox{TeV}~\lsim M_H^d \lsim 0.6$) TeV
for the parameter set I (II) if $\Delta M_{B_s}/\Delta M_{B_d}$
is equal to the experimental central value
 $35.05$ within an error of $1\%$.

\section{Conclusion}
We have considered a supersymmetric extension
of the SM
 based on the discrete $Q_6$ family symmetry,
 which has been recently proposed in 
  Refs.~\cite{Babu:2004tn,Kubo:2005ty,Kajiyama:2005rk}.
We have stressed  the one + two structure for each family;
one $Q_6$ singlet and one $Q_6$ doublet for
each family including the $SU(2)_L$ doublet Higgs fields.
We have found that it is possible to realize
the one + two structure  in a renormalizable way,
so that the Higgs sector becomes minimal 
and much simpler than that of the original  model of
 \cite{Babu:2004tn,Kubo:2005ty,Kajiyama:2005rk}.
 In this way the Higgs sector can be investigated with
 much less  assumptions.
It is  explicitly  shown that 
the SSB  parameters can be  fine tuned
 so as to make the heavy Higgs bosons much heavier than
$M_Z$ and at the same time to obtain
a desired size of spontaneous CP violation 
to reproduce the Kobayashi-Maskawa CP violating phase.

We have investigated the FCNC
 processes, especially those mediated 
 by heavy neutral Higgs bosons.
 Because of the $Q_6$ family symmetry,
 the number of the independent Yukawa couplings is smaller
 than that  of the observed quantities such as
 the CKM matrix
 and the quark masses. Therefore,
 the FCNCs can be parametrized only by
 the mixing angles and masses of  the Higgs fields:
 There are two angels and four mass parameters
 that enter into the FCNCs for a given $\tan\beta$;
 a set of three parameters for $\Delta M_D$
 and another set of three parameters for
 $ \Delta M_K$ and $\Delta M_{B_{d,s}}$.
 We have expressed the mass differences of the neutral
 mesons $\Delta M_K, \Delta M_D$ and $ \Delta M_{B_{d,s}}$
in terms of these parameters.
 
 Since the SM contributions to
$\Delta M_{B_s}$ and $\Delta M_{B_d}$ are well-controlled,
we haven taken into account  them to obtain the
constraints from $\Delta M_{B_s}$ and $\Delta M_{B_d}$.
That is, we have assumed that the extra contributions are allowed
only in a small window in which the SM values differ from the experimental values.
Allowed ranges in which the constraints are satisfied
are shown in various figures, where
$\Delta M_K, \Delta M_{B_s}$ and $\Delta M_{B_d}$
take values in the common parameter space.
We have also investigated
the ratio $\Delta M_{B_s}/\Delta M_{B_d}$
in the  region, in which all the constraints
from $\Delta M_{B_s}$ and $\Delta M_{B_d}$
are simultaneously
satisfied, and found that in a wide subregion
the ratio differs from
the experimental central value only
by less than 5\%. 
If we require that all the constraints
from $\Delta M_K, \Delta M_{B_s}$ and $\Delta M_{B_d}$
including 
the ratio $\Delta M_{B_s}/\Delta M_{B_d}$ are satisfied, 
we have found that
the heavy Higgs bosons should be heavier
than $\sim 1.5$ TeV.
If we relax the constraint from 
$\Delta M_K$ 
to $\Delta M_K^{\rm EXTRA}
< 2 \Delta M_K^{\rm exp}$
(because of the reason
that  non-perturbative contributions
suffer from large uncertainties),
the heavy Higgs bosons can be as light as
$\sim 0.4$ TeV, which is
 within the accessible range of LHC \cite{Buttar:2006zd}.

\vspace{0.5cm}
\noindent
{\large \bf Acknowledgments}\\
We would like to thank K.~Babu, Y.~Kajiyama,
H.~Okada and D.~Suematsu
for useful discussions. 
This work is supported by the Grants-in-Aid for Scientific Research 
from the Japan Society for the Promotion of Science
(\# 18540257).

\end{document}